\documentclass[fleqn,usenatbib]{mnras}

\usepackage[T1]{fontenc}

\DeclareRobustCommand{\VAN}[3]{#2}
\let\VANthebibliography\thebibliography
\def\thebibliography{\DeclareRobustCommand{\VAN}[3]{##3}\VANthebibliography}

\usepackage{graphicx}	
\usepackage{amsmath}	
\usepackage{amssymb}	

\usepackage{siunitx}
\usepackage{color}
\usepackage{xcolor}
\usepackage{tabularx}

\usepackage{ulem}

\usepackage{newtxtext,newtxmath}

\title[Chemodynamics of accreted gas in dwarf galaxies]{Spatially-resolved chemodynamics of the starburst dwarf galaxy CGCG\,007-025: Evidence for recent accretion of metal-poor gas}

\author[M. G. del Valle-Espinosa et al.]{
Macarena G. del Valle-Espinosa,$^{1}$\thanks{E-mail: macarena.garciavalle@ed.ac.uk}
Rub\'en S\'anchez-Janssen,$^{2,1}$
Ricardo Amor{\'i}n,$^{3,4}$
Vital Fern{\'a}ndez,$^{3,4}$ 
\newauthor
Jorge S{\'a}nchez Almeida,$^{5,6}$
Bego{\~n}a Garc{\'i}a Lorenzo, $^{5,6}$
Polychronis Papaderos,$^{7,8}$
\\
$^{1}$Institute for Astronomy, University of Edinburgh, Royal Observatory, Edinburgh EH9 3HJ, UK\\
$^{2}$UK Astronomy Technology Centre, Royal Observatory, Blackford Hill, Edinburgh EH9 3HJ, UK\\
$^{3}$Departamento de Astronom\'ia, Universidad de La Serena, Av. Juan Cisternas 1200 Norte, La Serena, Chile\\
$^{4}$Instituto de Investigaci\'on Multidisciplinar en Ciencia y Tecnolog\'ia, Universidad de La Serena, Ra\'ul Bitr\'an 1305, La Serena, Chile\\
$^{5}$Instituto de Astrof{\'i}sica de Canarias, E-38205 La Laguna, Tenerife, Spain \\
$^{6}$Departamento de Astrof{\'i}sica, Universidad de La Laguna, Tenerife, Spain \\ 
$^{7}$Centro de Astrof\'isica e Ci\^{e}ncias do Espaço, Universidade de Lisboa - OAL, Tapada da Ajuda, PT1349-018 Lisboa, Portugal\\
$^{8}$Instituto de Astrof\'{i}sica e Ci\^{e}ncias do Espaço - Centro de Astrof\'isica da Universidade do Porto, Rua das Estrelas, 4150-762 Porto, Portugal \\
}

\date{Accepted XXX. Received YYY; in original form ZZZ}

\pubyear{2022}

\begin{document}
\label{firstpage}
\pagerange{\pageref{firstpage}--\pageref{lastpage}}
\maketitle

\begin{abstract}

Nearby metal-poor starburst dwarf galaxies present a unique opportunity to probe the physics of high-density star formation with a detail and sensitivity unmatched by any observation of the high-z Universe. Here we present the first results from a chemodynamical study of the nearby, gas-rich starburst dwarf CGCG 007-025. We use VLT/MUSE integral field spectroscopy to characterise the properties of the star-forming (SF) gas, from its metal content to its kinematics. The star formation rate (SFR) surface density presents a clumpy distribution, with the brightest knot hosting a 5 Myr young, Wolf-Rayet (WR) population (revealed by the presence of the characteristic 5808\AA~ WR bump). The ionised gas kinematics are dominated by disordered motions. A superposition of a narrow  ($\sigma \approx$ 50 km s$^{-1}$), intermediate (150 km s$^{-1}$) and broad (1000 km s$^{-1}$) kinematic components are needed to model the emission line profiles in the brightest SF region, suggesting the presence of energetic outflows from massive stars. The gas-phase metallicity of the galaxy spans 0.6 dex and displays a strong anti-correlation with SFR surface density, dropping to 12+log(O/H) = 7.7 in the central SF knot. The spatially-resolved BPTs indicates the gas is being ionised purely by SF processes. Finally, the anti-correlation between the SFR and the gas metallicity points out to accretion of metal-poor gas as the origin of the recent off-centre starburst, in which the infalling material ignites the SF episode. 
\end{abstract}




\defcitealias{jr:vital}{F22}



\section{Introduction}
\label{intro}

\setcitestyle{notesep={ }}
Observations \citep[][]{jr:reviewMZR} and models of galactic chemical evolution \citep{jr:MZRma} indicate that massive galaxies tend to have higher gas-phase metallicities than low mass systems. The so-called mass-metallicity relation \citep[MZR, ][]{jr:tremonti2004} naturally emerges from the combination of secular and dynamical processes. First, the gas content in massive galaxies is in a more evolved stage and hence are chemically richer \citep{jr:tinsley&larson1978,jr:edmunds1990}. Second, low mass galaxies, thanks to their lower gravitational potential, are unable to retain most of its metals, which are lost through galaxy winds and outflows \citep[e.g., ][]{jr:chisholm2015,jr:tortora2022,jr:xu2022}. 
\setcitestyle{notesep={, }}
The MZR presents substantial scatter --specially and more severely in the low mass regime \citep{jr:curti}--  which seems to be associated with the star formation rate \citep[SFR, ][]{jr:massmetsfr,jr:laralopez2010}, so that for galaxies with the same stellar mass, the less metallic systems will have higher SFR.
The interplay between galaxy mass, SFR and metallicity is thought to arise as a result of the complex equilibrium between inflows, outflows, and mergers.

Theory and simulations predict that galaxies with high baryonic mass fractions ($f_{\text{gas}} > 0.3$) have a turbulent interstellar medium \citep[ISM, ][]{jr:HaywardHopkins}, characterised by clumpy structures and whose star formation histories (SFHs) are regulated by inflows and outflow of material \citep{jr:fire}. Although high gas mass  fractions were common among the high-z galaxy population, in the local Universe this almost exclusively occurs in low mass systems \citep{jr:bradford2015}. 
Thus, nearby dwarf starburst galaxies are considered the best candidates when looking for high-z analogues, sharing stellar masses, metallicities and specific SFRs with the average high-redshift galaxy \citep{jr:wiesz2011,jr:ML2011, jr:brammer2012,jr:garland2015,jr:sanchezalmeida2016,jr:izotov2021}. 

However, low mass galaxies reside in low mass haloes which are usually highly inefficient in capturing, retaining \citep{jr:dave2009,jr:christensen2018} and converting baryons into stars \citep{jr:geha2006,jr:kaufmann2007} due to their shallow potential wells.
\setcitestyle{notesep={ }}
In addition, the absence of stabilising structures --like bars or spiral arms-- in dwarf galaxies reduce the effectiveness of the starbursts. Alternatively, mergers and interactions between gas-rich galaxies with comparable masses can induce strong starbursting episodes \citep{jr:luo2014, jr:knapen2015, jr:pearson2019}. Simulations show that the tidal torques by interactions canalise large quantities of preexisting metal-poor gas from the outskirts towards the central regions \citep{jr:fire}. Moreover, the interaction-induced high gas pressures not only translate into higher star formation efficiencies, but also facilitate the formation of star clusters and super star cluster complexes \citep[see ][ for a review]{jr:adamo2020}.\setcitestyle{notesep={, }}
Hence, interactions between dwarf galaxies can effectively spark and sustain star formation for longer periods of time \citep{jr:patton2013,jr:moreno2021}. Although dwarf-dwarf interactions must be extremely abundant in the early Universe (due to the hierarchical nature of structure formation), the vast majority of dwarf galaxies in the Local Universe are satellites or isolated galaxies, with less than 5\% being part of a dwarf galaxy pair or a dwarf-only group \citep{jr:stierwalt2015,jr:besla2018}.

Cold accretion of gas from the intergalactic medium (IGM) is also predicted to produce star formation in dwarf galaxies, at high and low redshifts \citep[e.g., ][]{jr:dekelbirnboim2006,jr:dekel2009}.
Direct detections of gas inflowing into a galaxy are relatively difficult to observe \citep{jr:gasaccretion}. First, its absorption features are very weak, mainly because of its expected low density and low metallicity \citep[e.g., ][]{jr:fumagalli2011}. Second, its covering fraction may be very small \citep[less than 10\%, e.g., ][]{jr:fumagalli2011,jr:rubin2012}, requiring an alignment with the line of sight \citep{jr:ho2019}. Various efforts have been made to detect the pristine gas accreting into galaxies by using absorption systems \citep[e.g., ][]{jr:crighton2013,jr:zabl2019}. However, metal-poor gas accretion leaves some imprints in the characteristics of the galaxy that can be more easily detected. Bright, off-centre starburst clumps are typically seen as a result of intense star formation episodes induced by gas accretion. These off-centre starburst clumps have been detected vastly in intermediate and high redshift galaxies \citep{jr:elmegreen2013,jr:amorin2014,jr:amorin2015,jr:guo2015,jr:calabro2017}. As good analogues of these high redshift star-forming galaxies \citep{jr:papaderos2008,jr:papaderos2012}, local dwarf galaxies also show various signs of star formation in their outskirts. For example, \citet{jr:tadpoles} studied the bright star-forming regions of 7 tadpole galaxies \citep[also known as cometary, ][]{jr:loosethuan1986}, a particular type of object formed by an elongated diffuse component with a bright clump in one of the extremities. They found these peripheral star forming regions have a deficit in metallicity, and associated its origin to a recent event of gas accretion \citep[see also ][]{jr:BCDinhomogeneities,jr:sanchezalmeida2015}. 

In this vein, starburst dwarf galaxies (in which we can include blue compact dwarfs, BCDs; extremely metal poor galaxies, XMP; or HII galaxies) are the most promising local analogues for such chemically unevolved dwarf galaxies, since they tend not only to be metal-poor systems but also to have a high fraction of gas \citep[HI mass fraction above $30\%$, ][]{jr:amorin2009,jr:amorin2016,jr:filho2013,jr:filho2016,jr:trebitsch2017} as well as strong stellar feedback \citep{jr:fire}. Their strong bursts of star formation are usually on top of a much older stellar population, dominating the detected optical emission \citep{jr:loosethuan1986,jr:bergvallostlin2002,jr:amorin2007,jr:amorin2009}. Kinematical analyses of some BCDs indicate erratic velocity fields that point to dwarf galaxy mergers and/or in-falling of large gas clouds as the sources of their starburst \citep{jr:ostlin2001,jr:moiseev2015,jr:watts2016}. 

The development of Integral Field Spectroscopy (IFS) allowed not only to improve these studies but also to compare the properties of the SF regions with their surroundings. 
\citet{jr:cresci} reported inverted metallicity gradients in high redshift galaxies using near-IR IFS data: the gas-phase metallicity decreases towards the centre of the galaxies, coincident with the star forming regions. Similar findings are presented in \citet{jr:troncoso2014} for Lyman-break galaxies at $z\sim3.4$. This suggest very metal poor gas was accreted into the centre of the galaxies.
Using data from the MaNGA survey, \citet{jr:ALM} found some off-centred young star forming regions with an anomalously low-metallicity. They found these regions are more easily found in blue low-mass galaxies, and they concluded their origin was the accretion of metal-poor gas. \citet{jr:sanchezmenguiano2019} found using 700 MaNGA galaxies \citep{jr:MaNGAbundy2015} that local enhancements in star formation coincide with drops in gas phase metallicity.
Recently, the spatial metallicity variations ($\Delta$O/H $\approx$ 0.7 dex) seen in \citet{jr:fernandezarenas2022} were attributed to the accretion of pristine gas which fell into the system during merging stages.

In this paper  we use MUSE IFS observations to study the ionised gas properties of the dwarf galaxy CGCG 007-025. The paper is organised as follows. In Section \ref{sec:data}, we introduce the data and spectral fitting routine. In Section \ref{sec:Halpha}, the distribution and properties of the ionised hydrogen gas are presented. We obtain the chemical abundances and compute the ionisation maps in Section \ref{sec:pcg}, in addition with the spatially resolved BPT diagrams \citep[BPT, after ][]{jr:BPT1981}. Section \ref{sec:SFregions} is devoted to the individually detected star forming knots. In Section \ref{sec:discussion}, our main findings are discussed in the framework of the gas accretion scenario. In Section \ref{sec:conclusions}, we give a summary and present our conclusions.

\section{Data and methodology}
\label{sec:data}


\begin{figure*}     
    \begin{center}
        \includegraphics[trim=0 30 0 0,clip, width=0.98\textwidth]{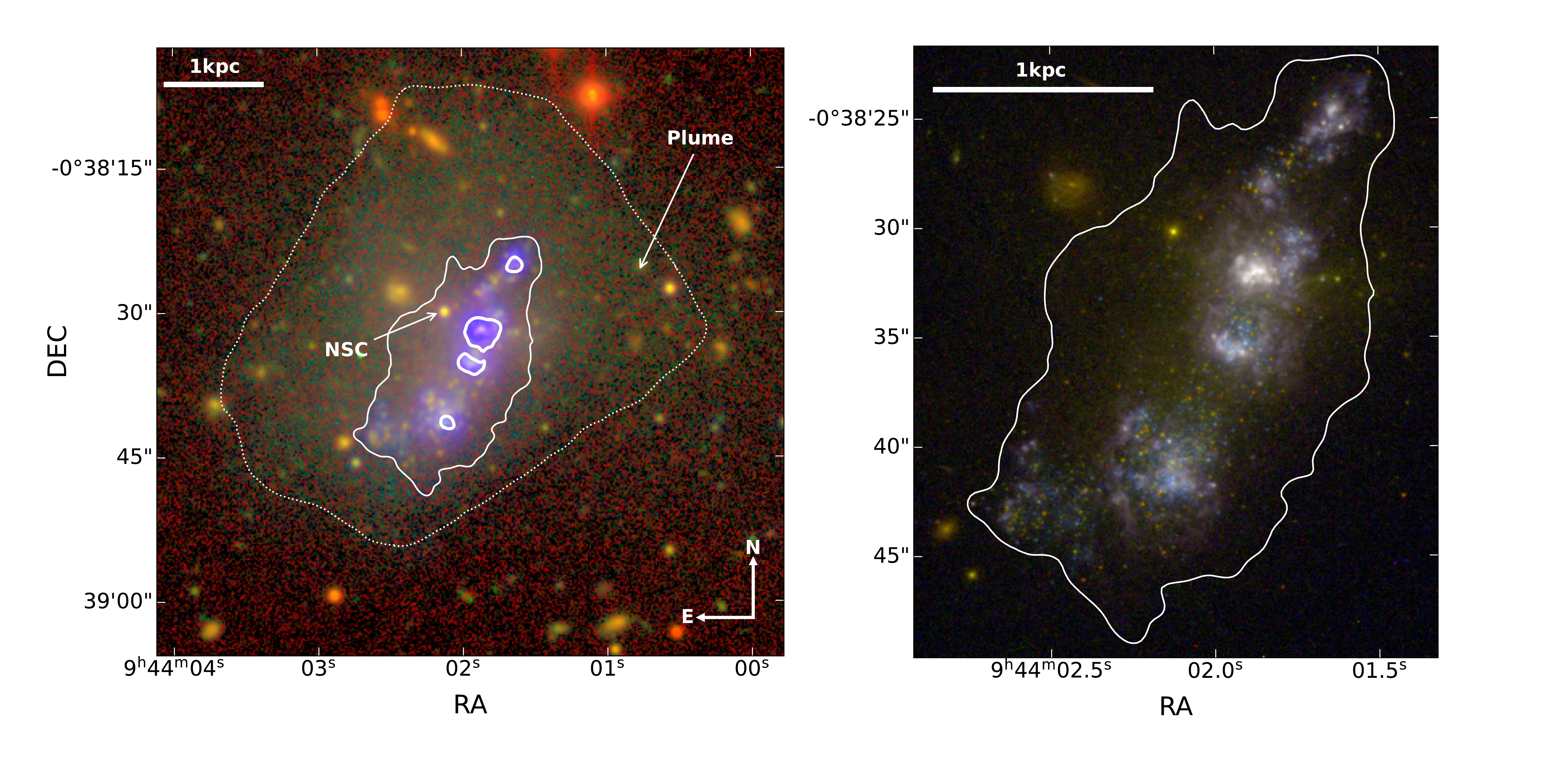}
    \end{center}
\caption{\textit{Left panel:} SUBARU/HSC $giy$ colour image of the starbursting galaxy CGCG 007-025, where \textit{g, i} and \textit{y} filters correspond to the colours blue, green and red, respectively. The host galaxy displays boxy isophotes and a prominent, central nuclear star cluster. A plume of low surface brightness material is also apparent towards the West, resulting from the interacting with its nearby dwarf companion, UGC\,5205. The thin-solid-white contour represent the extended H$\alpha$ emission detected with MUSE down to $\sim 1\times 10^{-16}$ erg/s/cm$^2$, while the dotted-white contour correspond to the isophote at $\mu_{i} = 26$ mag\,arcsec$^{-2}$ of the SUBARU/HSC image. Thick-solid-white contours delimit the four star-forming regions analysed in Sec. \ref{sec:SFregions}. \textit{Right panel:} Zoomed HST/WFC3 colour image performed by combining the F275W, F335W, F427W, F606W, F657N, F875W filters. The white-solid line is the same contour level as in the left panel.}
\label{fig:CGCG}                 
\end{figure*}

\subsection{Target description}

CGCG 007-025, from the \textit{Catalogue of Galaxies and Clusters of Galaxies} \citep[\citealt{jr:zwicky1968}, other cross-IDs are MCG +00-25-010 and SDSS J094401.86-003832.1, ][]{jr:cgcgMGC,jr:cgcgSDSS}, is known as a low metallicity starburst galaxy with extremely intense emission lines. At a distance of $d_L = 23 \pm 5$ Mpc \citep{jr:distanceCGCG}, CGCG 007-025 is a dwarf galaxy (${\rm M_{\star} = 1.2 \times 10^8 ~M_\odot}$, \citealt{jr:marasco_2022}) in interaction with another dwarf --UGC 5205-- at a projected separation of 8.3 kpc \citep{jr:zee_evolutionary_2000}. The interaction between them has triggered recent star formation episodes in both galaxies. 

This galaxy has been extensively studied as a prototypical nearby, metal-poor starburst \citep[][ among others]{jr:izotov&thuan2004,jr:zee&haynes2006,jr:senchyna_2017,jr:Senchyna2022ageCGCG,jr:bergCLASSY}, and as a local analogue to high-z low mass galaxies \citep{jr:brammer2012}. However, recent ground-based Subaru observations taken under the Hyper Suprime-Cam (HSC) Survey make more evident the presence of the underlying old component of CGCG 007-025 (Figure \ref{fig:CGCG}). In a forthcoming study (del Valle-Espinosa et al., in prep.) we present a detailed study of the host galaxy, which is characterised by old stellar populations and boxy isophotes. The photometric centre of the isophotes does not match the location of the brightest star forming region in the galaxy. In fact, as revealed by HST/WFC3 imaging and VLT/MUSE spectroscopy, the isophotes share a common centre at the location of a nuclear star cluster \citep[NSC, ][]{jr:rsjNSC2019}. CGCG 007-025 is therefore best classified as an early-type, nucleated dwarf spheroidal galaxy exhibiting an off-centre starburst, very much akin to the well-known class of BCD galaxies with prominent underlying hosts \citep{jr:amorin2007,jr:amorin2009}. Figure \ref{fig:CGCG} shows the HSC \textit{giy}-colour image, in which the underlying stellar component of the galaxy, as well as the NSC, can be observed. The over-plotted solid-white contour represent the H$\alpha$ emission measure in the available IFS data for this galaxy (see Sec \ref{sec:sfr} for the description). The dotted-white contour corresponds to the isophote at $\mu_{i} = 26$ mag\,arcsec$^{-2}$. In this paper, we aim to study the properties of the off-centre starburst episode of the galaxy, and we will present the characteristics of the underlying population in a companion paper.


\subsection{Observational data}

Our study is based on archival observations collected with the Multi Unit Spectroscopic Explorer \citep[MUSE, ][]{jr:MUSE} at the VLT (Very Large Telescope; ESO Paranal Observatory, Chile)\footnote{Program ID 0102.B0325}. Observations were carried out with the nominal Wide Field Mode (WFM-NOAO-N), corresponding to a field of view (FoV) of $1\arcmin \times 1\arcmin$ with a spatial sampling of 0.2\arcsec. The spectral coverage ranges from 4800 to 9300 \AA\ with a resolving power of $R\approx 2600$ around the H$\alpha$ line. The observing conditions yield a FWHM effective spatial resolution of 0.7\arcsec. Data were already reduced using the MUSE pipeline \citep{pr:MUSEpipeline}.

A few spectral pixels present NaN values around the peak of the H$\alpha$ emission line. The location of these spaxels coincides with the brightest SF region in the galaxy. Upon inspection of the spectra associated to these NaN values we discovered that the [\ion{O}{iii}]$\lambda$5007\AA/[\ion{O}{iii}]$\lambda$4959\AA~ flux ratio is below the theoretical emissivity ratio $(\approx 2.97)$. We associate this behaviour to saturation and non-linearity of the CCD. In Section \ref{sec:extcor} we describe the procedure follow to account for these effects.


\begin{figure*} 
    \begin{center}
        \includegraphics[width=\textwidth]{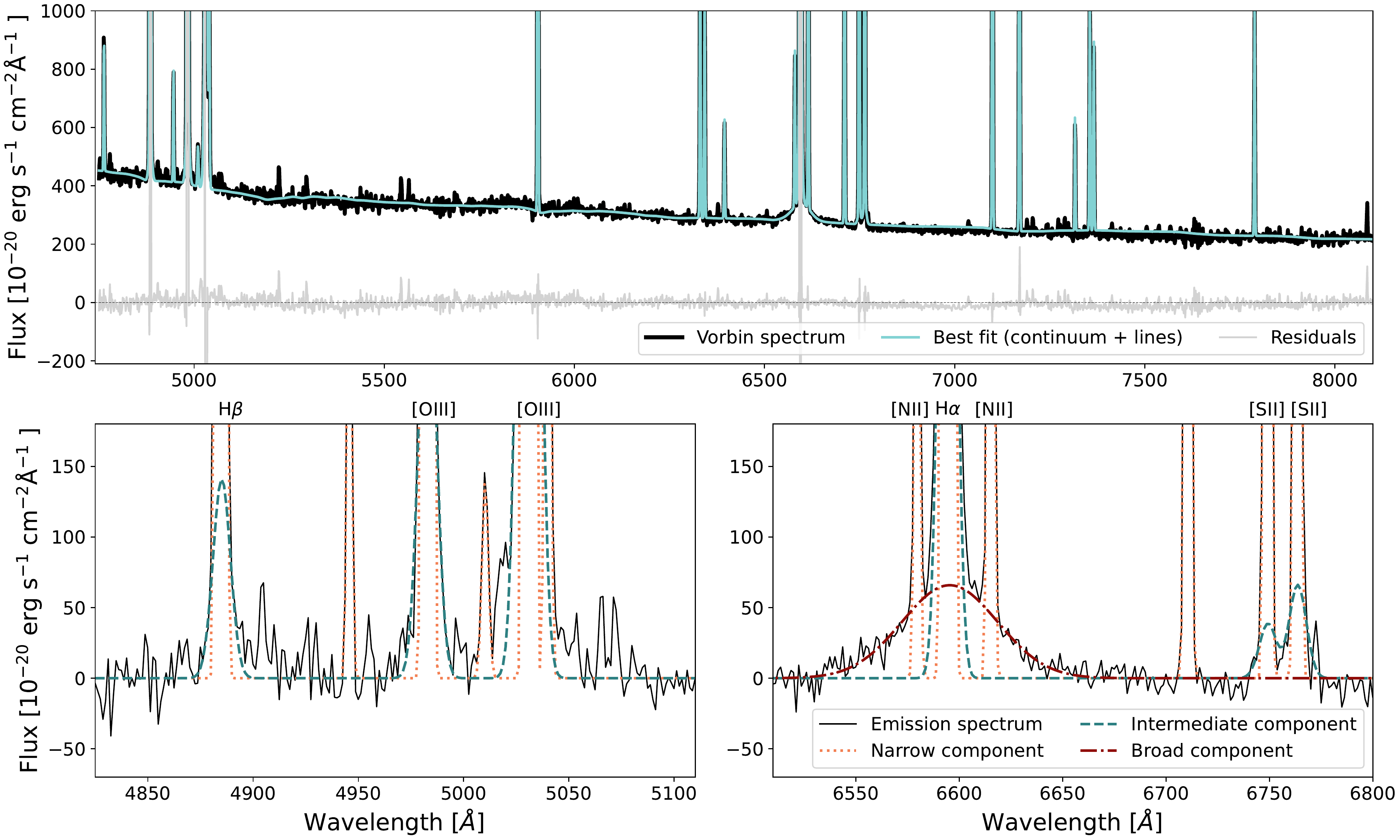}
    \end{center}
\caption{Top row: pPXF continuum + emission line fit (solid-cyan) for one binned spectrum of CGCG 007-025 (solid-black). The residuals of the fit are shown as a solid-grey line. Bottom row: zoom in showing the lines that required more than one component in the fit. The thin solid-black line represents the continuum-subtracted spectrum. In each panel, dotted-yellow lines represent the narrow component of each element and dashed-green lines correspond to intermediate components. The right panel also shows the broad component of H$\alpha$ (dash-dotted-brown).}
\label{fig:spectra}         
\end{figure*}

\subsection{Methods}
\label{sec:methods}

To extend the analysis of the data over the largest possible area voronoi tessellation was applied to the MUSE cube using the \textsc{VorBin} routine from \citet{pr:vorbin}. The tessellation wavelength definition was adapted according to the characteristic to study. For the detailed analysis of the H$\alpha$ emission (Sec. \ref{sec:Halpha}) we only keep spaxels with $S/N \geq 3$ in the 6560 - 6566 \AA ~rest-frame  wavelength range and target a minimum $S/N = 15$ for the resulting tessellated bin. This allowed us to reach the faintest nebular emission in the galaxy. When studying the ionisation and chemistry of the object (Sec. \ref{sec:pcg}) we performed the voronoi tessellation around the faintest set of lines we need for this purpose: the sulphur doublet [\ion{S}{ii}]$\lambda\lambda$6717,6732\AA. In this case, the previous $S/N$ cuts were defined over the range 6710 - 6740 \AA.

The procedure for the fitting routine starts by determining the stellar continuum. We mask every emission line and model the stellar continuum using \textsc{pPXF} \citep{pr:ppxf}, which finds the best fit for each spectrum-bin considering a linear combination of spectral templates. We use the single stellar populations (SSPs) templates from the \textsc{E-miles} library \citep{jr:emiles}, which cover the full MUSE spectral range.

Once we model the stellar continuum we subtract from the original MUSE datacube, creating a pure-gas cube. We split the gas cube in three spectral regions: from 4750 to 5200 \AA , gathering the H$\beta$ and [\ion{O}{iii}]$\lambda\lambda$4959,5007\AA~  emission lines; from 6400 to 6700 \AA , for the analysis of H$\alpha$ and [\ion{N}{ii}]$\lambda\lambda$6548,6584\AA; and from 6700 to 6800 \AA , which covers the emission of [\ion{S}{ii}]$\lambda\lambda$6717,6732\AA. We expect the nebular continuum to have little contribution to the overall line fluxes \citep{jr:byler2017}. Nevertheless, we add a one degree polynomial to model the continuum level in these three windows separately that should account for the smooth nebular continuum contribution. Due to the presence of broad wings in the base of the brightest emission lines, sometimes we use the sum of up to three gaussian profiles to model the overall emission line. 

Some of the species only require a narrow component for the fit (such as the nitrogen doublet), while the others show more complex kinematic components. For these cases, we distinguish between lines that may require one additional kinematic component (hereafter \textit{intermediate} component) and lines where even a third component is needed to fit the profile (the \textit{broad} component). The latter case is only relevant for the H$\alpha$ emission line, whereas an intermediate component is frequently necessary in H$\beta$ and in the oxygen and sulphur doublets. The fitting routine starts with narrow-only components and successively adds as many components as described above. 
We use the Bayesian Information Criterion (BIC) to decide how many components are required to model the data \citep{jr:BIC1978}. The model with the smallest BIC value is considered to provide the best description of the emission. Additionally, we only preserve kinematic components whose peak S/N is above 5. Lines in doublets share kinematic properties (i.e. red/blue-shifting and velocity dispersion), and their height ratios are fixed to the theoretical values when appropriate \citep[3 for the oxygen doublet and 2.95 for the nitrogen doublet;][]{jr:BOOKosterbrock}. The fits to the H$\alpha$ and H$\beta$ lines are fully independent, since they are fitted in different spectral windows and their ratios contain information on the obscuration by dust. Typical flux uncertainties for the narrow component are $\sim5\%$, while for the intermediate and broad components are $\sim20\%$. The velocity uncertainties are  are 0.1\%, 2.5\% and 10\% for the narrow, intermediate and broad components, respectively. Uncertainties in the velocity dispersion are 3.5\% for the narrow component while for the intermediate and broad components are $\sim20\%$. For the spaxels in which we require the presence of three kinematic components in the H$\alpha$ line, the narrow component is typically always dominant (97\% of the total line flux), with the intermediate and broad components representing a much smaller contribution (2\% and 1\%, respectively). Unless explicitly mentioned, the analyses in this study always refer to the narrow component. We correct the velocity dispersion from instrumental broadening using the derived values from \citet{jr:MUSEres}, which correspond to $\sim$50 km/s for the H$\alpha$ emission line.

Once all the lines are fitted and the number of components is defined, we correct all line fluxes of Milky Way extinction using the Cardelli attenuation law \citep{jr:MW_extinction}. At the location of CGCG 007-025, the adopted E(B-V) value for the Milky Way corresponds to 0.035 \citep{jr:green2018}.

Figure~\ref{fig:spectra} shows an example of the best model for a selected bin, where intermediate components are favoured for H$\alpha$, H$\beta$, [\ion{O}{iii}] and [\ion{S}{ii}] lines, and a broad component is also required for H$\alpha$. While we model all lines present in the spectra (see Table \ref{tab:emlist}), our analysis is focused on eight main lines: H$\beta$, [\ion{O}{iii}]$\lambda\lambda$4959,5007\AA, H$\alpha$, [\ion{N}{ii}]$\lambda\lambda$6548,6584\AA~ and [\ion{S}{ii}]$\lambda\lambda$6717,6732\AA. The rest of the lines present in Table \ref{tab:emlist} are fitted for a better constrain of the continuum and/or the additional kinematic components.


\subsection{Extinction correction}
\label{sec:extcor}

\citet{jr:vital} (hereafter \citetalias{jr:vital}) obtained the $c(H\beta)$ map of CGCG 007-025 using the Large Magellanic Cloud (LMC) reddening law by \citet{jr:LMC_gordon} with a $R_V = 3.1$. When the signal-to-noise was high enough, they computed the $c(H\beta)$ coefficient using H$\beta$ and the available Paschen lines, whereas when the signal-to-noise decreases, they used H$\beta$ and H$\alpha$ for the computation of the extinction. With this approach they can map $c(H\beta)$ in the brightest spaxels of the galaxy, where the H$\alpha$ and [\ion{O}{iii}]$\lambda$5007 lines were saturated. 
They found the central clump of CGCG 007-025 has a higher dust content than the rest of the galaxy, with a peak of extinction at $E(B-V)=0.35\pm0.04$ and dropping to $E(B-V)= 0.15\pm0.06$. When moving to the outskirts, a mean value of $E(B-V)= 0.08\pm0.08$ was found. For a more detailed description of the method and inferred extinction we refer the reader to \citetalias{jr:vital}. For the spaxels in which the $c(H\beta)$ is known, we use the values derived in \citetalias{jr:vital} to correct from intrinsic extinction. Otherwise, we corrected from extinction assuming the outskirts value $E(B-V) = 0.08$.

Finally, for the spaxels in which the H$\alpha$ and [\ion{O}{iii}]$\lambda$5007 lines were saturated, we substitute the modelled fluxes for the narrow component by the corresponding fluxes given by the theoretical  H$\alpha$/H$\beta$ and [\ion{O}{iii}]$\lambda$5007/[\ion{O}{iii}]$\lambda$4959 ratios, respectively \citep{jr:BOOKosterbrock}.

\section{The Star-forming Gas}
\label{sec:Halpha}

\begin{figure}
    \begin{center}
        \includegraphics[trim= 80 50 30 50,clip, width=0.48\textwidth]{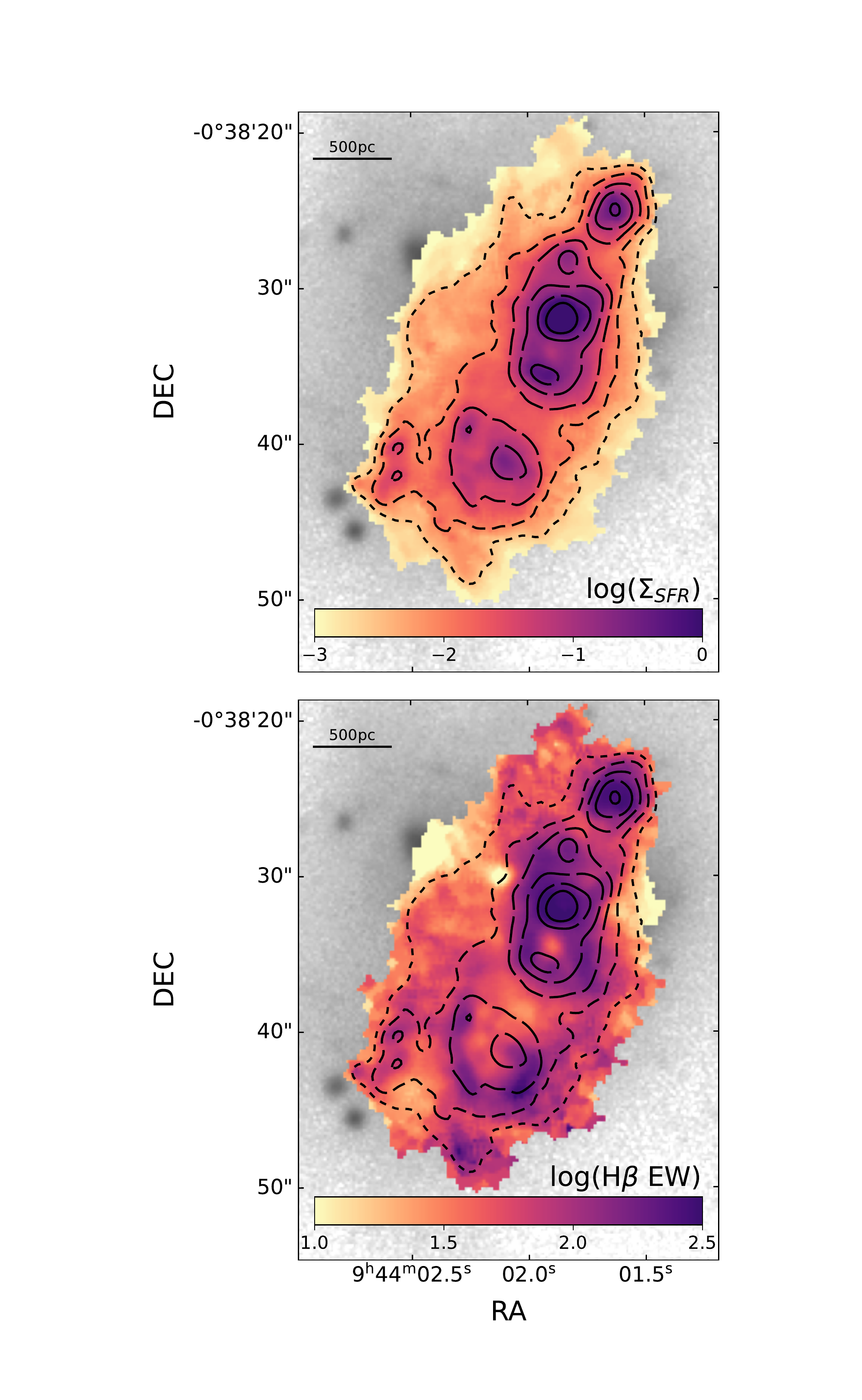}
    \end{center}
\caption{Maps of SFR surface density (top), in ${\rm M_{\odot}yr^{-1}kpc^{-2}}$, and EW(H$\beta$) (bottom), in \AA, for CGCG 007-025. Contour levels connect points with ${\rm log}(\Sigma_{\text{SFR}})$ of [-2.5, -2, -1.5, -1, -0.5, 0]. The background image in all displayed maps always corresponds to the SUBARU/HSC \textit{g} filter.}
\label{fig:EWandSFRmaps}
\end{figure}


\subsection{Star formation rate}
\label{sec:sfr}

One of the brightest emission lines widely used to estimate the star formation rate (SFR) in the literature is the H$\alpha$ recombination line, which traces the ionised gas in the surroundings of massive stars. Since H$\alpha$ is directly linked to the presence of massive stars younger than $\sim$10 Myr, the derived SFR represents the instantaneous value \citep{jr:SFRKennicutt,jr:SFRKenEvans}. Using a Kroupa initial mass function \citep{jr:kroupaIMF} and the \textsc{Starburst99} stellar models \citep{jr:stb99} at $1Z_\odot$, \citet{jr:SFRformula1} updated the SFR(H$\alpha$) calibration originally presented in \citet{jr:SFRKennicutt}, finding ${\rm SFR(H\alpha)[M_\odot yr^{-1}] = 5.37 \times 10^{-42} L(H\alpha)[erg~s^{-1}]}$. We convert the extinction-corrected H$\alpha$ fluxes into luminosities, and derive the SFR surface density (${\rm \Sigma_{SFR}}$) by dividing the SFR by the area of each bin (in ${\rm kpc^{2}}$) for all the spaxels. In the upper panel of Figure \ref{fig:EWandSFRmaps} we present the ${\rm \Sigma_{SFR}}$ map, with the different contours representing the $\log ({\rm \Sigma_{SFR}[M_{\odot}yr^{-1}kpc^{-2}]})$ levels at [-2.5, -2, -1.5, -1, -0.5, 0].
${\rm \Sigma_{SFR}}$ derived from the H$\alpha$ emission line can suffer from systematic uncertainties, mainly coming from the adopted dust correction and from variations in the predicted ionising flux of massive stars. However, \citet{jr:SFRCALIFA} studied the reliability of the ${\rm \Sigma_{SFR}}$ obtained from the dust-corrected H$\alpha$ surface brightness in 104 SF galaxies from the CALIFA survey. They compared this value with the hybrid  ${\rm \Sigma_{SFR}}$ derived using the FUV and 22$\mu m$ emission, finding a good agreement between them especially when ${\rm \Sigma_{SFR}(H\alpha) > 10^{-3} M_\odot yr^{-1}kpc^{-2}}$. 

We compute the global instantaneous SFR of the galaxy by integrating the ${\rm \Sigma_{SFR}}$ distribution, which yields a value of ${\rm \log(SFR/(M_\odot yr^{-1})) =  -0.67 \pm 0.09 }$. \citet{jr:marasco_2022} derived the total SFR for this galaxy combining the FUV emission from GALEX and the W4 (22 $\mu m$) from WISE. They found the total SFR of the galaxy to be ${\rm \log(SFR/(M_\odot yr^{-1})) = -0.64 \pm 0.13}$. \citet{jr:bergCLASSY} also compute the SFR of CGCG 007-025 by modelling the SED of the object using the FUV and NUV data from GALEX, and the \textit{ugriz} photometry from SDSS, corrected from aperture effects. Their modelling recovers a SFR value of ${\rm \log(SFR/(M_\odot yr^{-1})) = -0.78^{+0.19}_{-0.16}} $. Therefore, our estimation computed using the total H$\alpha$ luminosity is in good agreement with the values in the literature.


\subsection{H~\textsc{ii} gas mass}
\label{sec:massHII}

In order to calculate the total mass of ionised gas present in the galaxy we follow the procedure in \citet[][see also \citet{jr:carniani2015}, \citet{jr:olmogarcia2017}]{jr:mHII}. 
The mass of the ionised gas is directly proportional to the ratio of the H$\alpha$ luminosity, $L_{\text{H}\alpha}$, and the electron density, $n_e$, following:

\begin{equation}
    M_{\text{\ion{H}{II}}} = \dfrac{\mu ~m_\text{H}}{h~ \nu_{\text{H}\alpha}~\alpha^{\text{eff}}_{\text{H}\alpha}}\times\dfrac{L_{\text{H}\alpha}}{n_e}
\end{equation}

\noindent
where $\mu$ is the atomic weight (assumed to be 1.0), $m_\text{H}$ is the hydrogen mass, $h$ is Planck's constant, $\nu_{\text{H}\alpha}$ is the H$\alpha$ frequency and $\alpha^{\scriptscriptstyle\text{eff}}_{\scriptscriptstyle\text{H}\alpha}$ \normalsize is the coefficient for H$\alpha$ in the Case B recombination. The latter coefficient depends on the electron temperature of H$\alpha$, so we need to derive first the electron density and temperature in order to apply the formula.

\citetalias{jr:vital} inferred the electron density and two electron temperatures (for the high and low ionisation species) of individual spaxels where the [\ion{S}{iii}]$\lambda$6312\AA~ line was reliable enough to apply the direct method. Briefly, the [\ion{S}{ii}]$\lambda\lambda$6717,6732\AA~ ratio was used to obtain the electron densities, the ratio between [\ion{S}{iii}]$\lambda\lambda$6312,9069\AA~ for the low ionisation temperature, and the electron temperature of the high ionisation species was derived using the empirical relation from \citet{jr:hagele2006}.
The low ionisation species were found to have a temperature of $(1.5 \pm 0.1) \times 10^4$ K, which corresponds to a value of $ \alpha^{\scriptscriptstyle\text{eff}}_{\scriptscriptstyle\text{H}\alpha} = (8.0 \pm 0.7) \times 10^{-14}$ cm$^3$ s$^{-1}$  \citep{jr:alphaeff}. 
The electron densities of the brightest star forming clumps in CGCG 007-025 were also obtained in \citetalias{jr:vital}. The density of the brightest star-forming knot is as high as $n_e = 380 \pm 60 ~ {\rm cm^{-3}}$, with lower densities in the other three knots reaching $n_e = 90 \pm 50 ~ {\rm cm^{-3}}$. We further assume that this value applies to the outer parts of the galaxy. 

With the electron density values on hand as well as the H$\alpha$ luminosity, we create the ionised gas mass surface density map. By integrating this map, we estimate the total amount of ionised gas is $M_{\text{\ion{H}{II}}} = (1.3 \pm 0.9)\times 10^6 M_\odot$. A comparison of this value with the stellar mass estimated by \citet{jr:marasco_2022} for the galaxy reveals the ionised gas mass is around $1\%$ of the stellar mass. 

\begin{figure*}
    \begin{center}
        \includegraphics[trim=20 70 30 30,clip, width=0.98\textwidth]{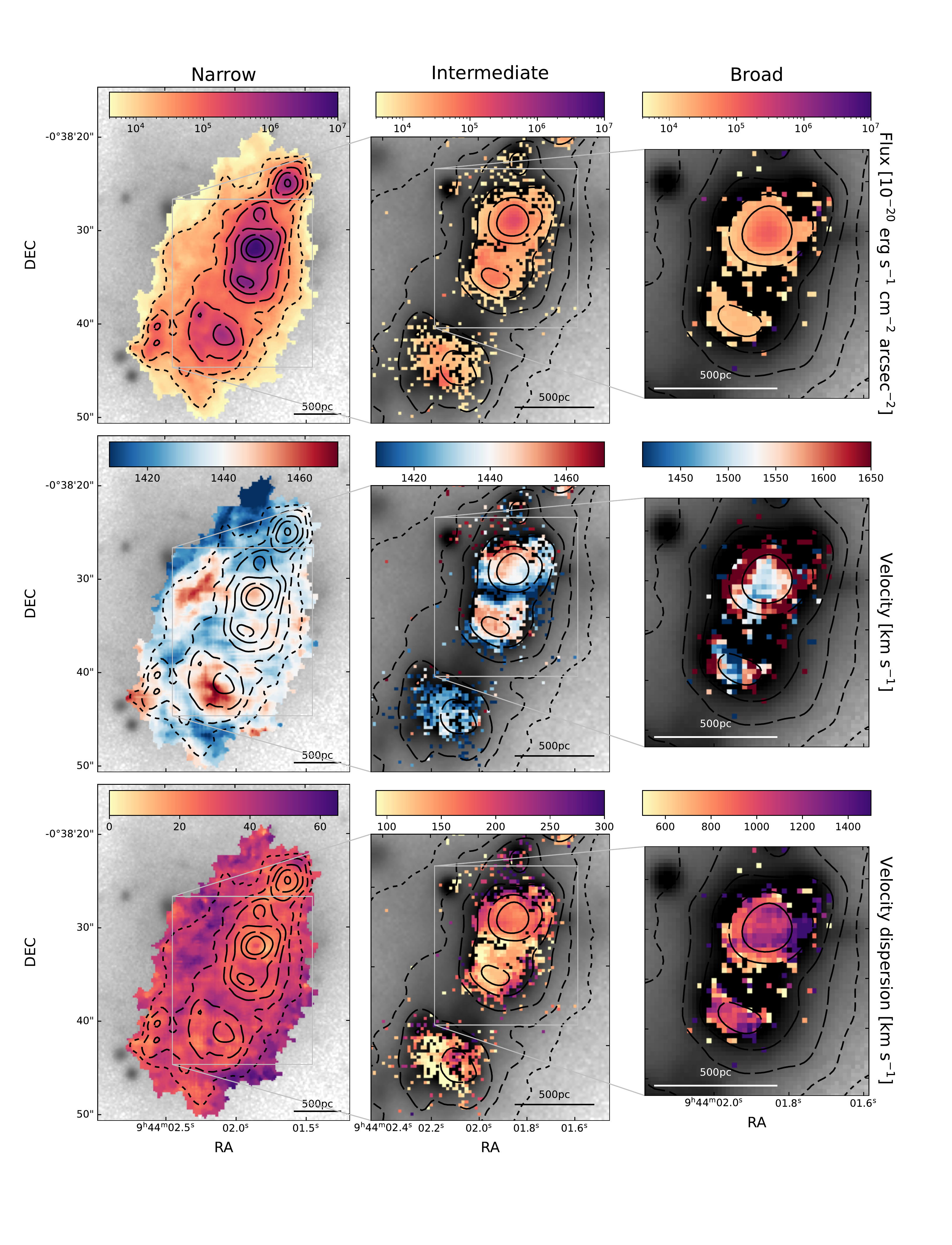}
    \end{center}
\caption{From top to bottom: flux, velocity field and velocity dispersion --corrected from instrumental broadening-- for the three H$\alpha$ kinematic components: narrow (left), intermediate(centre) and broad (right). Black contours connect points with same $\Sigma_{\text{SFR}}$, as in Fig. \ref{fig:EWandSFRmaps}. Note the different spatial scale used for the three components. We find unordered motions in the gas velocity fields. Blue and red colours in the velocity maps correspond to blueshift and redshift. The velocity of the broad component is systematically redder than the velocity of the other components.}
\label{fig:Halphamaps}
\end{figure*}


\subsection{EW map}
\label{sec:ew}

The equivalent widths (EWs) of the strongest hydrogen recombination lines correlate closely with the age of the most recent burst \citep{jr:dottori&bica1981,jr:EWHbIntro_other,jr:EWTerlevich,jr:EWHbIntro}. On the one hand, the most massive and hottest stars produce a nebular component that will drop rapidly with time due to their short lifetimes. On the other hand, the underlying absorption will slowly decrease because of the significant contribution from low mass stars. By studying the ratio of these two features we can measure the ratio between the young and old population, giving an estimation of the age of the most recent burst. 

Among the different hydrogen recombination lines, the EW of the H$\beta$ line is a proven age indicator up to $t\sim10$ Myr independently of the star formation history adopted \citep{jr:EWconstSFH}. Although the EW of the hydrogen recombination lines is affected by the metallicity of the population, several authors \citep[among others]{jr:O7V/Otime,jr:EWHbIntro,jr:bpassV2.1} have characterised its dependence using different stellar libraries and evolutionary synthesis codes. Thanks to these studies, the EW(H$\beta$) is a reliable proxy to determine the age of recent starburst.

We build the EW map using the narrow component of the H$\beta$ recombination line since the continuum around this line is better determined than in the case of H$\alpha$. As a sanity check we also compute the H$\alpha$ EW map, which leads to the same distribution. Figure \ref{fig:EWandSFRmaps} displays the EW(H$\beta$) map, in logarithm scale, in the bottom panel. EW(H$\beta$) peaks at the location of the star-forming clumps, with values EW(H$\beta$) $\geq 100$ \AA. The position of the NSC is very clear on the map, where the EW drops to values below 1 \AA. 

A comparison of the EW(H$\beta$) values in the brightest  knot with those from the Binary Population and Spectral Synthesis \citep[\textsc{bpass}, ][]{jr:bpassV2.1,jr:bpassV2.2} indicates the starburst is very young, with an upper limit to the age of $t \leq 10$ Myr.


\subsection{Kinematics}
\label{sec:kin}

Figure~\ref{fig:Halphamaps} shows the maps for the flux (top row), velocity (middle row) and velocity dispersion (bottom row) of the three kinematic components of the H$\alpha$ line: narrow (left column), intermediate (middle column) and broad (right column). The contours displayed in all the maps are the same $\Sigma_{\text{SFR}}$ levels as in Figure~\ref{fig:EWandSFRmaps}.

The flux distribution map of the narrow component expands 3.1 kpc in the NW-SE direction and 1.6 kpc in the NE-SW. When compared with the area encompassed by the isophote at $\mu_{i} = 26$ mag\,arcsec$^{-2}$ (Figure \ref{fig:CGCG}), the ionised gas covers around 45\% of the galaxy. The ionised gas emission is concentrated in four main star-forming regions with an additional diffuse extended component. As highlighted in Figure \ref{fig:CGCG}, the morphology of the ionised gas is significantly different in all four main star-forming regions. From the S-E region to the central clump the gas morphology changes from being noticeably dispersed (as are the stars) to being significantly more concentrated (with the stars still embedded). Moreover, this change in the gas morphology seems to be linked to the age of the regions: the S-E region has the lowest EW(H$\beta$) –the oldest SF region– while the central region has the largest one –the youngest SF region–. In this scenario, the arc has an intermediate value of EW(H$\beta$) when compared with these other two regions, with its stars more concentrated than in the S-E region and its gas forming a cavity around them. The intermediate component is present in the four main star forming knots, whereas the broad is just in the brightest one.

The velocity field of the narrow component --middle left panel of Figure \ref{fig:Halphamaps}-- does not exhibit any coherent kinematic structure, with instrumental-limited velocity dispersion values almost constant all across the galaxy --lower left panel of Figure \ref{fig:Halphamaps}--. Our findings for the narrow component are in agreement with the maps present in \citet{jr:marasco_2022}. In the southwest of the galaxy, the velocity map seems to be chaotic, with small patches with opposite velocity values. Across these patches, the velocity dispersion is slightly higher than at the position of the main bursts. Another region with higher velocity dispersion is located at the east side of the gas distribution, almost coincident with the position of the NSC. We cannot detect a rotational profile for the gas component in GCGC 0007-025 due to its highly perturbed velocity field.

The velocity field of the intermediate component --middle central panel of Figure \ref{fig:Halphamaps}--  presents, in the brightest SF clump, a velocity pattern which is compatible with a bipolar structure: a local velocity maximum and minimum are observed along the NE-SW direction, with relative velocities of 50 km s$^{-1}$. Moreover, the internal velocities of the gas in this region are higher than the others --note the higher velocity dispersion values in the lower central panel of Figure \ref{fig:Halphamaps}--.

On the central clump of CGCG 007-025 we uncover the presence of a broad component (right column of Figure \ref{fig:Halphamaps}). The gas associated with this component exhibits substantial internal motions, with velocity dispersions exceeding $\sigma \approx 1000$ km s$^{-1}$ (lower right panel of Figure \ref{fig:Halphamaps}). The velocity field (middle right panel of Figure \ref{fig:Halphamaps}) is compatible with a dusty expanding bubble, where its receding parts are not observed because of the high extinction of the region. We are not able to detect this broad component in the H$\beta$ emission line. This could be related to the addition of two effects: (1) the high extinction of the central clump, and (2) the low S/N we would expect for this line (for the H$\alpha$ broad component S/N $\sim10$). On the other hand, the narrow components of [\ion{O}{iii}]$\lambda$5007\AA~ and H$\alpha$ have similar S/N ($\sim 100$ for both lines), but no broad component is detected in [\ion{O}{iii}]$\lambda$5007\AA. A similar behaviour in low metallicity galaxies has being found in other galaxies \citep{jr:Izotov2010,jr:Cann2020}. This absence of broad emission in the forbidden lines could be explained if the broad lines are originated in a gas whose density is well above the critical density of the [\ion{O}{iii}], i.e. $n_e \sim 10^6~ {\rm cm^{-3}}$. 


\section{Physical properties of the ionised gas}
\label{sec:pcg}
\subsection{Chemistry}
\label{sec:chm}

In \citetalias{jr:vital} we not only computed the electron density and the electron temperatures (as briefly explained in Sec. \ref{sec:massHII}) but also the ionic and total abundances of the gas for the spaxels in which [\ion{S}{iii}]$\lambda$6312\AA~ was reliably measured. However, and since the [\ion{O}{iii}]$\lambda$4363\AA~ and [\ion{O}{ii}]$\lambda\lambda$3726,3729\AA~ lines fall out of the MUSE spectral range, we are not able to use the $T_e$-direct method to derive the oxygen abundance and the nitrogen-to-oxygen ratio in diffuse component of the gas distribution, where the [\ion{S}{iii}] lines are not detected. To infer the chemical properties of the gas in these regions we make use of the \textsc{HII-CHI-mistry}\footnote{\url{http://home.iaa.csic.es/~epm/HII-CHI-mistry.html}} \citep[v5.22, ][]{pr:hiichim,pr:hiichimv5.22}, a code that compares the emission line fluxes with a grid of photoionisation models and producs results that are compatible with the direct method. The ionising SED corresponds to a \textsc{bpass} model of a single instantaneous burst with an age of 1 Myr. The grid considers values for the ionisation parameter of $-4.0 <$ log(U) $< -1.5$ in steps of 0.25 dex. The metal abundances follow the Oxygen one, with 12+log(O/H) ranging from 7.1 to 9.1 in steps of 0.1 dex. This grid assumes 17 different values of the ratio log(N/O) in the range -2.0 to 0.0 in steps of 0.125 dex. 

\begin{figure*}
    \begin{center}
        \includegraphics[trim= 20 0 40 0,clip, width=1\textwidth]{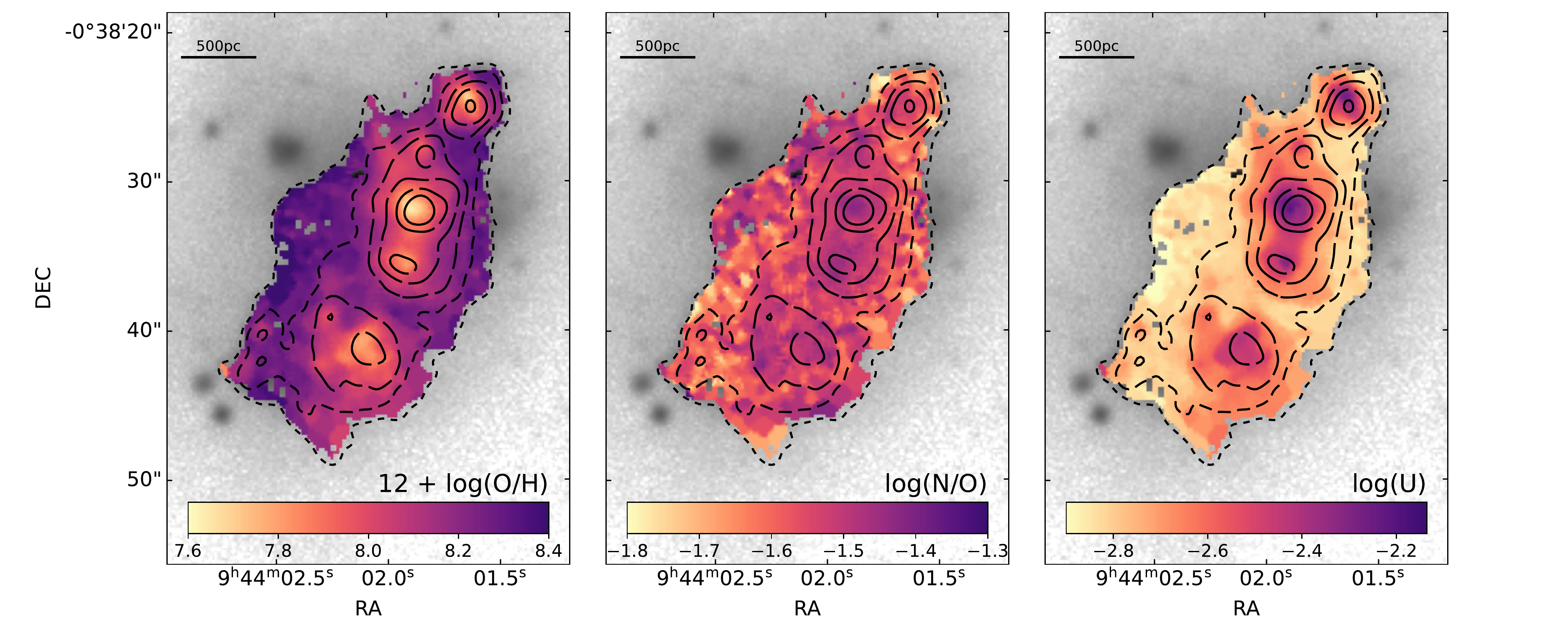}
    \end{center}
\caption{Oxygen abundance (left), nitrogen-to-oxygen ratio (centre) and ionisation parameter (right) maps of CGCG 007-025. Contour levels connect points with same $\Sigma_{\text{SFR}}$, as explained in Figure \ref{fig:EWandSFRmaps}. }
\label{fig:chemMaps}
\end{figure*}

Figure~\ref{fig:chemMaps} displays the maps for the three quantities computed by \textsc{HII-CHI-mistry}: 12+log(O/H), log(N/O) and log(U). The oxygen abundance distribution of the galaxy reaches its minimum in the central star forming knot, with a median value of 12+log(O/H) = 7.71 $\pm$ 0.09 , and increases towards the outskirts of the galaxy until 12+log(O/H) = 8.27 $\pm$ 0.10. Remarkably, the oxygen abundance also drops to a median value of 12+log(O/H) = 7.88 $\pm$ 0.09 in the other three star forming knots. When looking at the spatial distribution of the nitrogen-to-oxygen ratio, it remains essentially flat across the entire galaxy at a median value of N/O = -1.52 $\pm$ 0.21. The ionisation parameter follows the opposite distribution as the oxygen abundance, reaching maximal values where the star forming clumps are located (log(U) = -2.34 $\pm$ 0.08), and decreasing to log(U) = -2.82 $\pm$ 0.08 in the outskirts.

We can compare our results derived with \textsc{HII-CHI-mistry} with our direct-method estimations \citepalias{jr:vital} for the brightest spaxels in the galaxy. It is important to note that in \citetalias{jr:vital} we used the flux of eight transitions --[\ion{Ar}{III}], [\ion{Ar}{IV}], [\ion{N}{II}], [\ion{O}{II}], [\ion{O}{III}], [\ion{S}{II}], [\ion{S}{III}] and \ion{He}{I}-- to compute the chemistry of the galaxy, while for the method presented here we only use the strong lines [\ion{O}{III}], [\ion{S}{II}] and [\ion{N}{II}]. In \citetalias{jr:vital} we obtain a mean oxygen abundance of 12+log(O/H) = 7.88 $\pm$ 0.11. For the same spaxels, the gas-phase metallicity derived with \textsc{HII-CHI-mistry} is 12+log(O/H) = 7.90 $\pm$ 0.10. Both values are in good agreement. When looking at the N/O ratio, the direct-method returned log(N/O) = -1.84 $\pm$ 0.12, while the mean value derived in this paper is  log(N/O) = -1.50 $\pm$ 0.08. As explained in \citetalias{jr:vital} (left panel of their Figure 8), this difference is due to the degeneracy between log(N/O) and the flux of [\ion{N}{ii}]$\lambda$6584\AA~ when using photoionisation models: at 2$<$log(N/O)$<$ -1.25, negligible variations of the [\ion{N}{ii}]$\lambda$6584\AA~ flux result in huge log(N/O) fluctuations.

\begin{figure*} 
    \begin{center}
        \includegraphics[trim= 10 0 40 0,clip, width=1\textwidth]{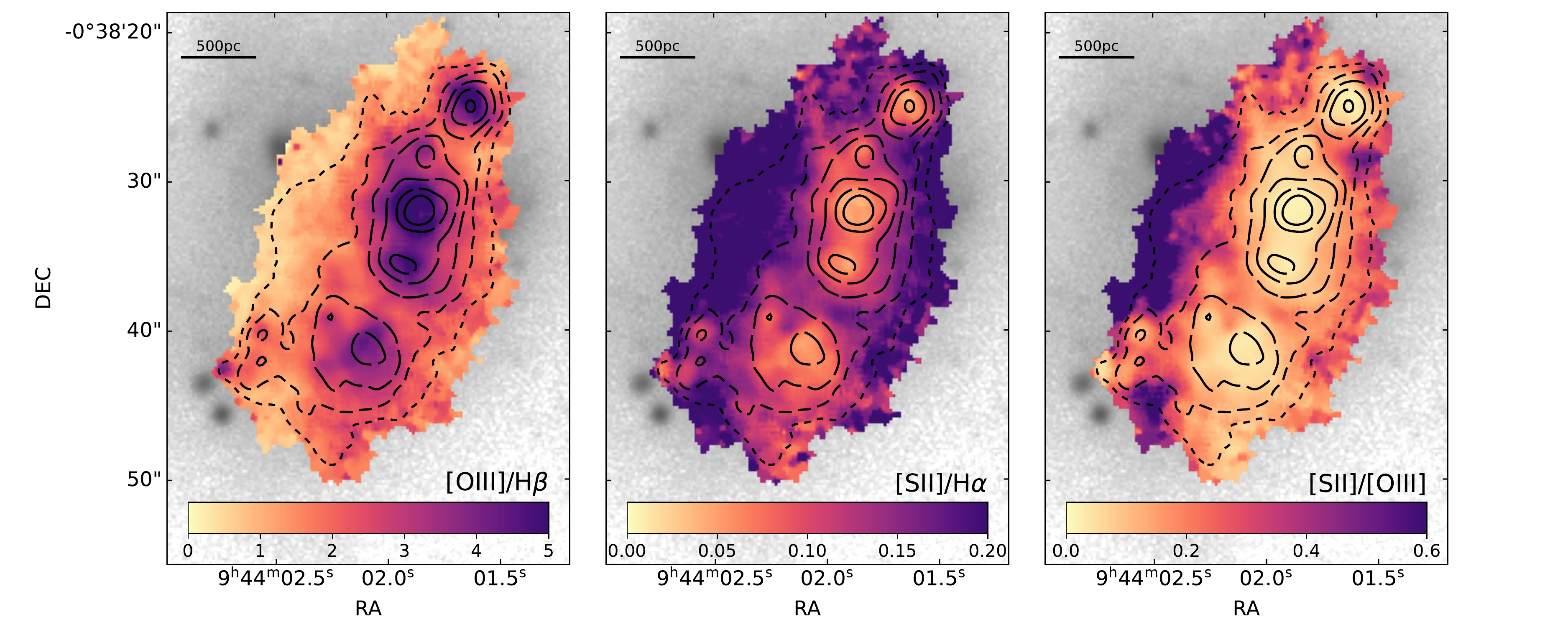}
    \end{center}
\caption{[\ion{O}{iii}]/H$\beta$ (left), [\ion{S}{ii}]/H$\alpha$ (centre) and [\ion{S}{ii}]/[\ion{O}{iii}] (right) emission lines ratio maps of CGCG 007-025. Contour levels are explained in Figure \ref{fig:EWandSFRmaps}. Peaks in [\ion{O}{iii}]/H$\beta$ are in the same position as lower [\ion{S}{ii}]/H$\alpha$ values, a typical behaviour of star forming regions. The ratio of [\ion{S}{ii}]/[\ion{O}{iii}] is lower than 0.5 all across the galaxy.}
\label{fig:IonisationMaps}
\end{figure*}

\subsection{Ionising mechanisms}
\label{sec:ionmaps}

The different excitation and ionisation mechanisms in the nebula are disentangled by studying different emission line ratios \citep{jr:BOOKdopita,jr:BOOKosterbrock}. Among the most common ratios we have access to [\ion{O}{iii}]/H$\beta$ for the high ionisation region and [\ion{N}{ii}]/H$\alpha$, [\ion{S}{ii}]/H$\alpha$ and [\ion{O}{i}]/H$\alpha$ for the low ionisation. In Figure \ref{fig:IonisationMaps} we display the maps of the [\ion{O}{iii}]/H$\beta$ (left) and [\ion{S}{ii}]/H$\alpha$ (middle) line ratios since they have the highest S/N, with the $\Sigma_{\text{SFR}}$ contours overplotted. 

The [\ion{O}{iii}]/H$\beta$ distribution peaks in the positions where the SF knots are located, and decreases towards the edges of the gas distribution. This distribution is expected when a recent star formation event has occurred, since [\ion{O}{iii}] is generated by high ionising sources like O-type stars. On the other hand, the distribution of the low ionisation line ratios present the opposite behaviour, with the minimums located at the position of the SF knots. This behaviour is also expected when the ionising source produces large quantities of UV photons, which also is the case for O-type stars.

\begin{figure}
    \centering
    \includegraphics[trim=0 65 0 20,clip, width=0.48\textwidth]{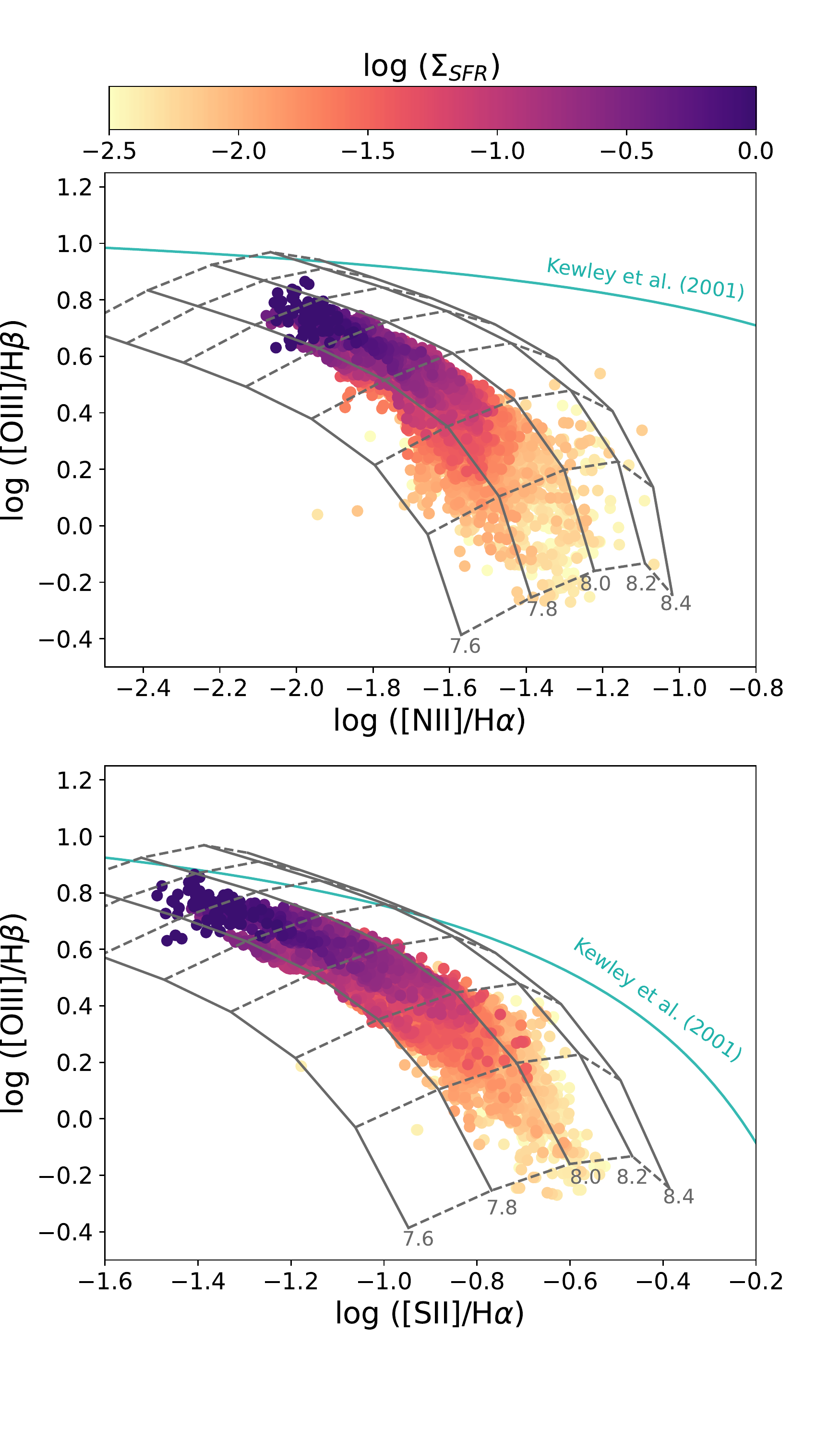}
    \caption{[\ion{O}{iii}]/H$\beta$ versus [\ion{N}{ii}]/H$\alpha$ (top) and versus [\ion{S}{ii}]/H$\alpha$ BPT diagrams colour-coded by the ${\rm \log (\Sigma_{SFR})}$. The \citet{jr:kewley} lines are displayed in both panels as reference of the solar behaviour. The \textsc{bpass} grids used to derive the chemistry of the object are overplotted, with the different metallicities annotated.}
    \label{fig:BPTs}
\end{figure}

The structure of an ionisiation-bounded SF cloud can be described by two ionisation regimes: the innermost region is where highly-ionised species reside, whereas the outer parts are dominated by low ionisation species. By studying the relative emission of two ions with different ionisation potential, we can disentangle the ionisation structure of our SF clumps. The right panel of Fig. \ref{fig:IonisationMaps} displays the [\ion{S}{ii}]/[\ion{O}{iii}] ratio, where [\ion{S}{ii}] is our low ionisation element and [\ion{O}{iii}] represents the highly ionised one. For an ionisation-bounded nebula the inner regions correspond to small values of [\ion{S}{ii}]/[\ion{O}{iii}], whereas the outer parts should reach values larger than one \citep{jr:pellegrini}.

The brightest SF knot in CGCG 007-025 has an average [\ion{S}{ii}]/[\ion{O}{iii}] value of 0.03, indicating the entire region is highly ionised. None of the 4 main star forming clumps present an [\ion{S}{ii}]/[\ion{O}{iii}] distribution expected for a classical ionised bounded nebula. This suggest the four SF clumps in the galaxy are optically thin, allowing the LyC photons to escape and ionise the surrounding gas \citep{jr:pellegrini,jr:bik_opticaldepth}.

Our emission-line ratio maps allow us to explore spatially resolved diagnostic diagrams \citep[the so called BPTs, ][]{jr:BPT1981,jr:BPT1987}. By comparing two emission line ratios of species with different ionisation levels, BPTs manifest the origin of the emission lines since the position of the spaxels reflects the gas ionisation state and source of ionisation. Using a set of photoionisation models, \citet{jr:kewley} conclude that, if the emission is coming purely from starbursting episodes, the points should always fall below an empirical limit. On the other hand, if other ionisation mechanisms are involved in the generation of the emission lines --such as AGN or shocks-- the points will move above this empirical limit. The position of this boundary in the BPT diagram depends mainly on the metallicity of the population and the ionisation parameter. 

In Figure \ref{fig:BPTs} we display two optical BPTs: [\ion{O}{iii}]/H$\beta$ versus [\ion{N}{ii}]/H$\alpha$ and [\ion{O}{iii}]/H$\beta$ versus [\ion{S}{ii}]/H$\alpha$ shown on the top and bottom panels, respectively. For reference, the \citet{jr:kewley} empirical lines are displayed in both plots. We remark that the \citet{jr:kewley} dividing lines are too conservative for a starburst galaxy with a metallicity as low as that for CGCG 007-025. In order to account for the metallicity deviation of CGCG 007-025 with respect to the solar value, we overplotted in both panels of Figure \ref{fig:BPTs} the theoretical grid used to derive the chemistry of the gas in Sec. \ref{sec:chm}. We limit the grid to the values obtained, so that log(N/O) = -1.5 and the oxygen abundance ranges from 7.6 to 8.4, while the ionisation parameter spans from -3.0 to -2.0. As shown in both diagnostic plots, the line ratios recovered for the narrow component can be entirely explained by star formation activity.


\section{Spectroscopy of individual star-forming clumps}
\label{sec:SFregions}

\begin{figure}
    \centering
    \includegraphics[trim=0 40 0 50,clip, width=0.49\textwidth]{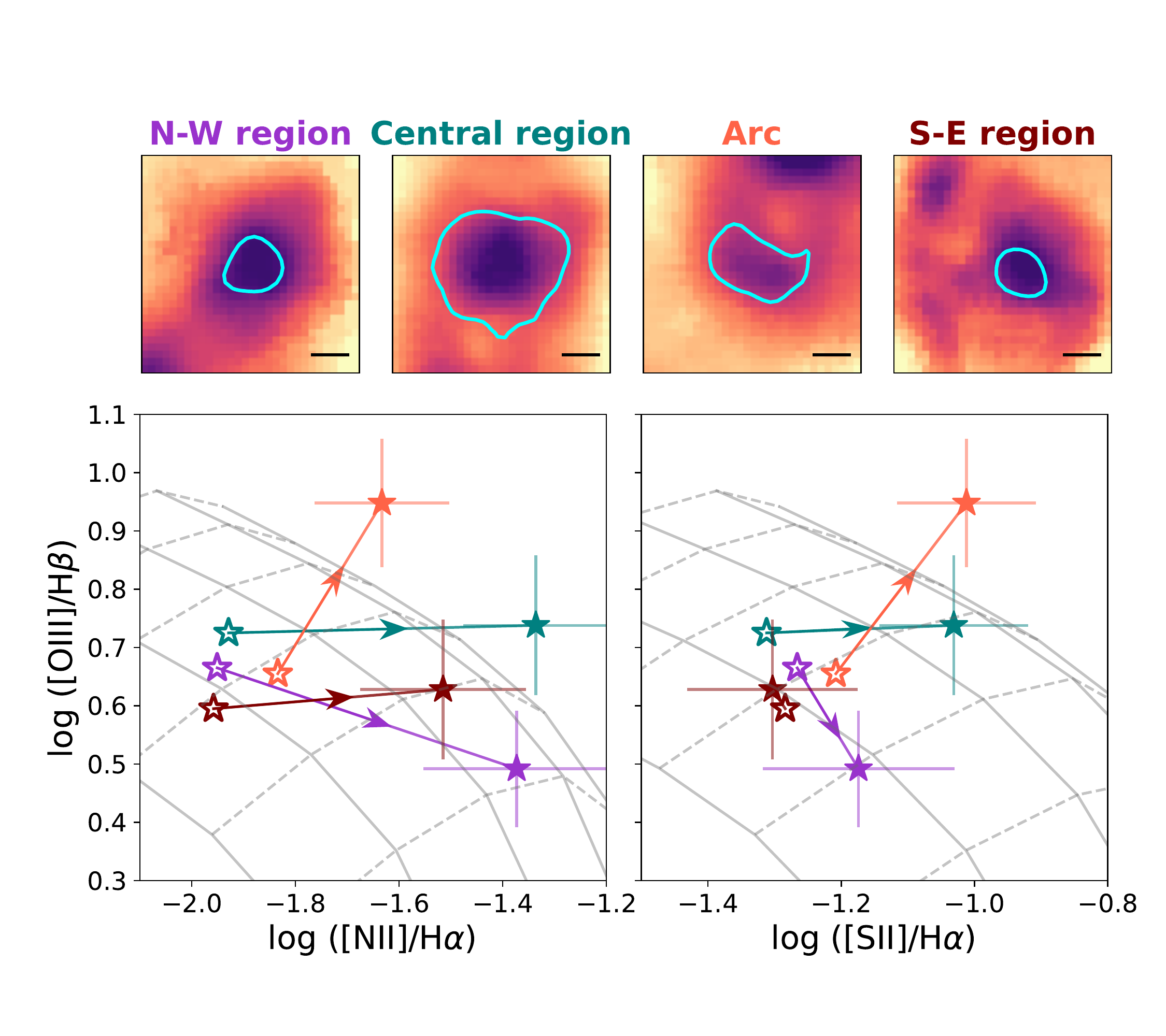}
    \caption{\textit{Top row}: Zoom into the H$\alpha$ flux distribution for the SF knots identify in \citetalias{jr:vital}. The cyan contours delimit each of the SF knots and correspond to a H$\alpha$ flux level of ${\rm 1 \times 10^{-17} erg~ s^{-1}cm^{-2}}$. The black line in the right corner of each panel represents a 100 pc scale. \textit{Bottom row}: BPT diagrams for the SF knots. Open stars correspond to the narrow component whereas filled stars stand for the intermediate component. Points from the same region are connected with a line. Uncertainties in the narrow component are smaller than the symbol size. The \textsc{bpass} photoionisation grids are overplotted in grey. Note the intermediate component move to the upper-right part of the diagnostic diagrams, leaving the parameter space that is explained by SF processes. }
    \label{fig:SFregions}
\end{figure}

\begin{center}
\begin{table*}
    \begin{tabular}{cccccccccc}\hline
    Clump & Distance to NSC & Area & EW(H$\beta$) & Velocity & log(L(H$\alpha$)) & log(SFR) & 12+log(O/H) & log(N/O) & log(U) \\
                &  kpc & kpc$^2$ & \AA & km s$^{-1}$ yr & erg s$^{-1}$ & ${\rm M_\odot~ yr^{-1}}$ & & & \\ \hline
    N-W region &  0.95 $\pm$ 0.03 & 0.02 & 283 & 1426.6 $\pm$ 0.2 & 39.01 $\pm$ 0.01 & -2.26 $\pm$ 0.01 & 7.91 $\pm$ 0.04 & -1.53 $\pm$ 0.05 & -2.48 $\pm$ 0.01 \\
    Central region & 0.48 $\pm$ 0.03 & 0.11 & 290 & 1443.8 $\pm$ 0.2 & 40.10 $\pm$ 0.01 & -1.17 $\pm$ 0.01 & 7.93 $\pm$ 0.04 & -1.46 $\pm$ 0.06 & -2.41 $\pm$ 0.02 \\
    Arc & 0.59 $\pm$ 0.03 & 0.05 & 124 & 1440.4 $\pm$ 0.2 & 39.36 $\pm$ 0.01 & -1.91 $\pm$ 0.01 & 7.94 $\pm$ 0.05 & -1.48 $\pm$ 0.06 & -2.56 $\pm$ 0.03\\
    S-E region & 1.12 $\pm$ 0.03 & 0.02 & 85 & 1451.8 $\pm$ 0.4 & 38.74 $\pm$ 0.01 & -2.53 $\pm$ 0.01 & 7.91 $\pm$ 0.06 & -1.53 $\pm$ 0.05 & -2.48 $\pm$ 0.01 \\
   
    \end{tabular}
    \caption{Derived parameters of the individual SF clumps present in the galaxy. }
    \label{tab:SFregions}
\end{table*}
\end{center}

In \citetalias{jr:vital}, we identified the four main SF knots of CGCG 007-025.
These regions were delimited at the  99.5 of percentile the flux distribution of all the spaxels in the datacube for the [\ion{S}{iii}]$\lambda$6312\AA~ emission line, the faintest line needed for the direct method. This percentile corresponds to a flux of ${\rm 9.3\times 10^{-20} erg ~s^{-1}cm^{-2}}$, implying all the spaxels have a flux of at least ${\rm 1 \times 10^{-17} erg~ s^{-1}cm^{-2}}$ for the H$\alpha$ emission. We add the spaxels with flux values above this H$\alpha$ flux limit to get the integrated spectrum of each SF clump.
Figure \ref{fig:SFregions} zooms into the H$\alpha$ flux distribution of each individual clump. The names displayed on top of the zoom images are first adopted in  \citetalias{jr:vital}. Note that the so-called \textit{central region} is our brightest SF clump and the closest to the NSC, even though it does not overlap with the NSC. With the data from Subaru/HSC it is evident that this region is not at the photometric centre of the galaxy, but we keep the name for consistency. The cyan contours mark the extent of the main SF clumps.

We analysed the spectra of the four clumps in the same way as explained in Sec. \ref{sec:methods}: we estimate the stellar continuum using \textsc{pPXF} and fit the different emission lines using the same multi-component gaussian modelling. From the emission line modelling we obtain the EW(H$\beta$), the velocity, the total H$\alpha$ luminosity and the SFR for each individual region. We also compute the integrated chemistry using \textsc{HII-CHI-mistry} with the same photoionisation models as before. The derived characteristics of the four SF knots are collected in Table \ref{tab:SFregions}. 

We found the chemical properties of the SF knots are similar among them, with no O/H or N/O differences from clump to clump. However, the values of EW(H$\beta$) increase from the southeast region to the northwest one. Given that the EW(H$\beta$) traces the age of the burst, with higher values of EW(H$\beta$) meaning younger stellar populations, the star forming episodes of the galaxy seem to have a delay between them. 

The brightest SF clump --central region in Table \ref{tab:SFregions}-- has spectrophotometric data from SDSS \citep{jr:cgcgSDSS}. \citet{jr:bergCLASSY} fit the {ugriz} SDSS photometry of the galaxy, complemented with FUV and NUV data from GALEX, and obtained a SFR value of ${\rm \log(SFR/(M_\odot~ yr^{-1})) = -1.49^{+0.21}_{-0.15}}$ for this region, in agreement with our value from the extinction-corrected H$\alpha$ emission.

The integrated spectra of the knots allow to better detect the intermediate components of [\ion{O}{iii}], [\ion{S}{ii}] and [\ion{N}{ii}]. The position in the BPTs of both the narrow and intermediate components of the lines are shown in the bottom row of Figure \ref{fig:SFregions}. The location of the narrow components are represented with open stars, while the filled stars stand for the intermediate components. The two components from the same region are connected by arrowed lines. As in Fig. \ref{fig:BPTs}, the theoretical \textsc{bpass} grids used to derive the chemistry of the galaxy are displayed in both panels. Note the open stars fall in the same position in the grid as the individual spaxels. The chemistry of the intermediate component is consistent within 2$\sigma$ with its narrow component chemistry, except in the \textit{arc} region. The intermediate component in the \textit{arc} moves out of the grid towards higher line ratios, leaving the regime in which SF purely explains the origin of the emission and some other mechanisms —like shocks— are needed to explain such excitation. 

A detailed visual inspection of the spectrum for the brightest complex of the galaxy (Figure \ref{fig:WR}) reveales the presence of numerous high excitation lines, such as [\ion{Fe}{vi}] or \ion{He}{ii}, as well as the presence of the Wolf-Rayet (WR) red bump. A more detailed analysis of this feature is presented in the next section.

\section{Discussion}
\label{sec:discussion}


\subsection{The ionising sources}
\label{sec:WR}

All the emission lines across the galaxy suggest they are generated by strong star formation. As the EW(H$\beta$) values suggest, the age of the burst is no older than $t=10$ Myr. We now turn our attention to the nature of the ionising sources. 

\begin{figure}
    \begin{center}
        \includegraphics[trim=0 10 0 0, clip, width=0.48\textwidth]{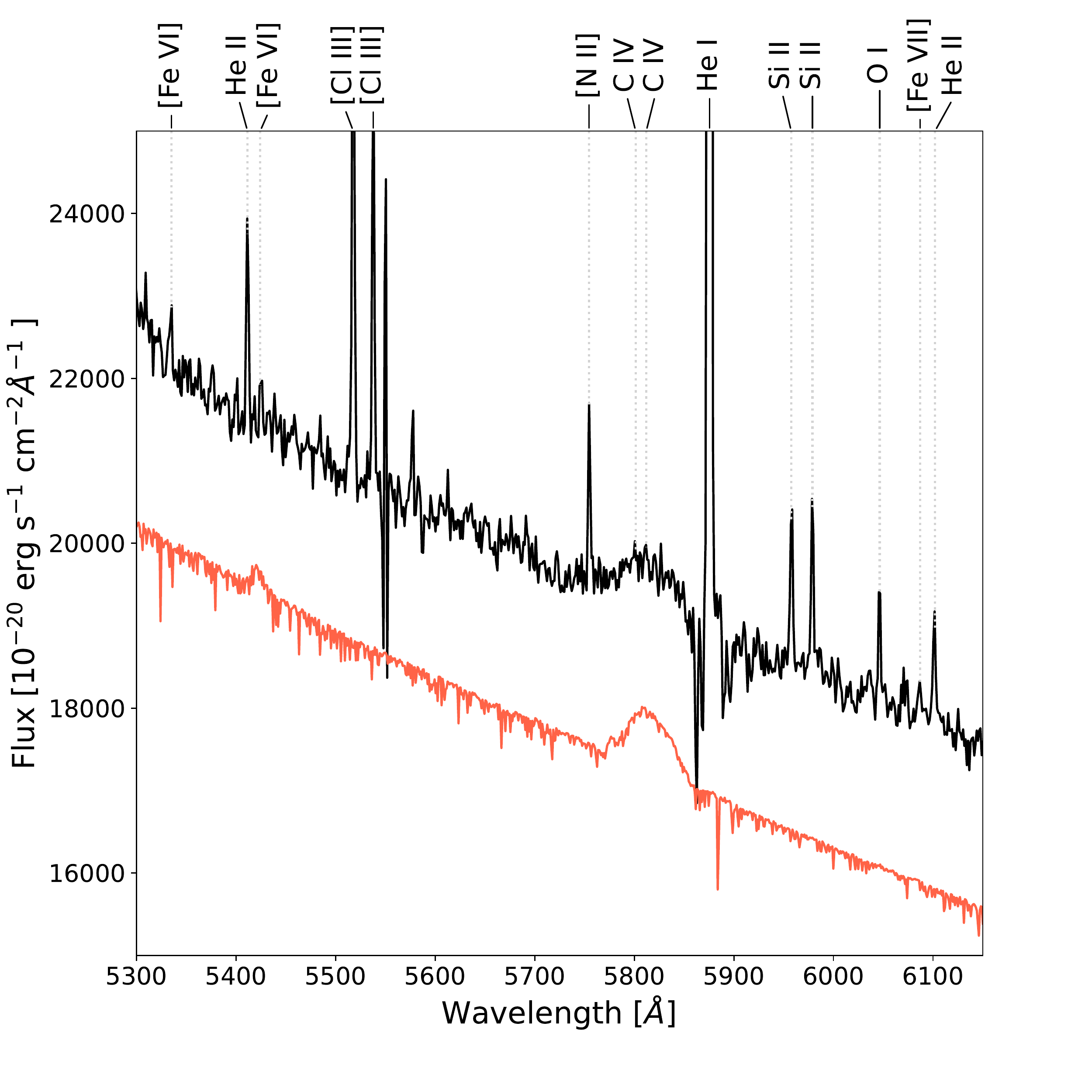}
    \end{center}
\caption{Detailed spectrum of the brightest SF knot in CGCG 007-025. The best fit to the \textsc{bpass} models is shown in orange, offset for visualisation purposes. High excitation lines, such as [\ion{Fe}{vi}] or \ion{He}{ii}, are also annotated with dotted lines.}
\label{fig:WR}
\end{figure}

The integrated spectrum of the brightest SF complex of the galaxy presents a broad emission feature at 5808 \AA~which is typically associated with the presence of WR stars. WR stars are a late evolutionary stage of massive stars in which the strong stellar winds expel the outer layers of the star \citep[e.g., ][]{jr:crowther2007}, exposing the innermost regions and reaching effective temperatures as high as 100,000 K \citep{jr:smith2002}. 
For single starburst episodes, the WR features appear between 3 to 5 Myr of the initial burst \citep{jr:stb99,jr:smith2002}. In the optical part of the spectrum the two most prominent WR features are broad bumps which originate from the emission of \ion{He}{ii} at 4686 \AA --the so called \textit{blue bump}-- and the emission of \ion{C}{iv} at 5808 \AA -- known as the \textit{red bump}-- \citep{jr:WNWC,jr:smith2016}. These broad emissions are generated by two different types of WR stars: nitrogen-rich WR (WN) are responsible for the blue bump whereas the red bump is generated by carbon-rich WR (WC) stars \citep{jr:WNWC}.
In Figure \ref{fig:WR}, we zoom in the spectrum of the brightest clump around the WR red bump region (black solid line). The different emission lines are annotated with dotted lines. In order to determine the presence of WR in MaNGA galaxies, \citet{jr:WRsignificance} define the significance, $\sigma_{\rm bump}$, of the WR bump as 
\begin{equation}
    \sigma_{\rm bump} = \dfrac{f_{\rm bump}}{f_{\rm rms}}
\end{equation}
where $f_{\rm bump}$ is the continuum-subtracted flux of the WR bump and $f_{\rm rms}$ is the noise of the spectrum around the baseline calculated in two windows at both sides of the red bump: 5660-5710\AA~ and 5900-5950\AA. Using the same formula in the collapsed brightest knot, the 5808 \AA ~bump is detected with a significance of $\sigma_{\rm bump} = 3.6$. 

Based on the gas-phase metallicity derive, we assume the metallicity of the stars which ionised the gas should be $1/10 Z_\odot$ \citep{jr:bpassV2.1}. At this metallicity, the presence of WR stars could be explained with models that incorporate binary populations. We use the \textsc{bpass} models in order to reproduce this feature. We fit the region displayed in Figure \ref{fig:WR} assuming the extinction follows the LMC \citep{jr:LMC_gordon} and recover the best fit corresponds to an age of t = 5.1 $\pm$ 1.2 Myr. \citet{jr:Senchyna2022ageCGCG} used deep UV spectroscopic data for the same region of the galaxy to derive an age of ${\rm \log(t/yr) = 6.7^{+0.5}_{-0.2}}$, in agreement with our age estimation. The spectrum of this population is displayed in Figure \ref{fig:WR} with a red solid line, with a vertical offset for better visualisation.

WR stars, as well as O-type stars, deposit a large amount of high energy photons into the surrounding \ion{H}{ii} region. In an optically thick nebula, all the ionising photons emitted by these stars will be absorbed. However, in an optically thin nebula, a fraction of these photons is able to reach the surrounding gas, ionising the interstellar medium (ISM). The ratio between the expected and the observed H$\alpha$ luminosity give us information about the fraction of photons escaping the ionising region. This escape fraction not only affects the available energy budget but also can probe the stars in the \ion{H}{ii} region are the responsible of the ionisation of the ISM.
The \textsc{bpass} models provide the expected H$\alpha$ luminosity in the case that all the ionising photons are absorbed by the nebula \citep{jr:bpassV2.2}. For our population, with an age of 5 Myr and 1/10 $Z_\odot$, the expected luminosity is $\log(L({\rm H\alpha_{exp}})) = 40.51 \pm 0.01$. The observed H$\alpha$ luminosity of the brightest complex is $\log(L({\rm H\alpha_{obs}})) = 40.23 \pm 0.02$. This implies that around 50\% of the total ionising photons are absorbed, so that the total escape fraction of the central clump is $f_{\rm esc} = 0.5$. Following a similar methodology, \citet{jr:dellabruna2021} inferred $0.2 < f_{\rm esc} < 0.8$ for typical \ion{H}{II} regions in NGC 7793, in agreement with our estimation for CGCG 007-025.

\begin{figure*}
    \begin{center}
        \includegraphics[trim=10 10 0 0, clip, width=0.98\textwidth]{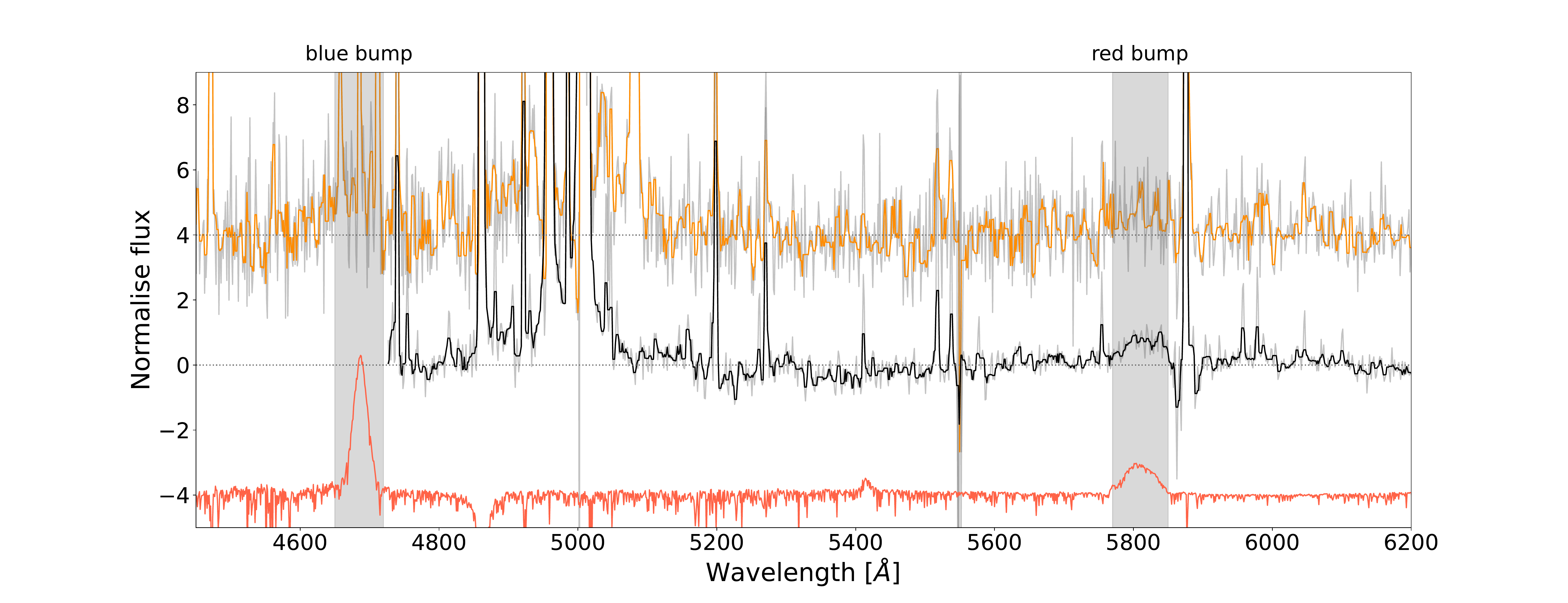}
    \end{center}
\caption{Comparison between the SDSS (orange) and MUSE (black) spectra using the same aperture. The red line represents the \textsc{bpass} best fit model shown in Fig. \ref{fig:WR}. Both WR bumps are highlighted in grey. The presence of the red bump is clear in the MUSE spectrum, but its limited wavelength range does not cover the position of the blue bump. The SDSS spectrum covers both WR bumps, but its lower S/N does not allow a reliable detection (although both highlighted regions are systematically above the continuum).}
\label{fig:WRSDSSMUSE} 
\end{figure*}

\subsubsection{Estimation of the number of WR stars}

The detection of the WR red bump allows us to estimate the total number of WC stars present in this complex. \citet{jr:WRwithMetallicities} computed an expression of the luminosity of one WC star depending on the oxygen abundance measured. Considering our region has 12+log(O/H) = 7.9, the luminosity of the red bump generated by one WC star is $L_{{\rm WC}} = 1.56 \times 10^{36} {\rm erg ~ s^{-1}}$. Since the measured bump has an extinction-corrected luminosity of $L_{{\rm red~bump}} = (3.61 \pm 0.56) \times 10^{37} {\rm erg ~ s^{-1}}$, the number of WC stars in the region is $N({\rm WC}) = 23 \pm 4$. To obtain the total number of WR stars (WC+WN) in the region we need first to estimate the number of WN stars. As mentioned above, WN stars generate a broad bump emission at 4686\AA, a region of the optical spectrum not accessible with the MUSE datacube but available in the SDSS spectrum. Despite covering a larger optical wavelength range, the WR bumps have not been reported in the SDSS spectrum. In Figure \ref{fig:WRSDSSMUSE}, we compare the SDSS spectrum of the galaxy (orange solid) with the MUSE spectrum (black solid) of the central knot, using the same aperture as in SDSS (i.e., 3\arcsec). Both spectra have been smoothed with a median filter to an effective resolution of 3 \AA , with the original spectra for each case in light grey. In red, the \textsc{bpass} best fit model to the MUSE spectrum is displayed, as in Fig. \ref{fig:WR}. The location of the two WR bumps at 4680 \AA~ and 5800 \AA is highlighted in grey. The detection of the red bump in the MUSE spectrum is clear, but much less so in the SDSS spectrum. However, the flux in the two bump regions always is above the continuum level in the SDSS spectrum. We measure a low detection  significance of ${\rm \sigma_{blue~bump}=1.3}$ and ${\rm \sigma_{red~bump}=1.2}$. Knowing from the MUSE data that the red bump is real, we model both bumps  in the SDSS spectrum with Gaussian components and infer a flux ratio $f_{{\rm blue~bump}}/f_{{\rm red~bump}} = 2.2 \pm 0.7$.

\citet{jr:WRwithMetallicities} also derive a metallicity-dependent relation for the luminosity of one WN, which corresponds to $L_{{\rm WN}} = 9.85 \times 10^{35} {\rm erg ~ s^{-1}}$ at the metallicity of this region. Using our measured value for the red bump in the MUSE spectrum, in addition with the ratio between the bumps derived from the SDSS spectrum, we estimate the number of WN stars as $N({\rm WN}) = 80 \pm 25$. The ratio of WC to WN stars then corresponds to ${\rm WC/WN} = 0.29 \pm 0.09$ and the estimated total number of WR in this region of the galaxy is ${\rm N(WR)} = 103 \pm 26$. With this value, we can perform a tentative estimation of the ratio of WR stars over O-type stars, $N({\rm WR})/N({\rm O})$. The number of O stars is related to the luminosity of the H$\beta$ line. For reference, a single O7V star produces a H$\beta$ luminosity of $L({\rm H\beta}) = 4.76\times10^{36} {\rm erg~ s^{-1}}$ \citep{jr:VaccaConti1992}. O7V stars represent a fraction of the total population of O stars, $\eta_0$, which depends on the shape of the IMF, upper mass limit and age of the burst. \citet{jr:O7V/Otime} calculate the time evolution of $\eta_0(t)$ for a Salpeter IMF with an upper mass limit of $120 M_\odot$, taking into account the evolution of the O star population as well as the emergence of WR stars. For a population with an age of 5 Myr, this value is $\eta_0\approx 0.25$. The total H$\beta$ luminosity of the region is $L({\rm H\beta}) = (6.04\pm 0.06)\times10^{39} {\rm erg~ s^{-1}}$, which yields to a total of $N({\rm O}) = 5080 \pm 50$. The ratio $N({\rm WR})/N({\rm O})$ derived for this region is $N({\rm WR})/N({\rm O}) = (20 \pm 6) \times 10^{-3}$. As reference, \citet{jr:senchyna_2017} used the SDSS spectrum of the central clump to analyse the possible WR population. Although they do not detect any broad feature in any of the two WR bump locations, they give a tentative value of the ratio $N({\rm WR})/N({\rm O})$ to be lower than 0.02, compatible with our findings. However, the \textsc{BPASS} models predict a total of 63 $\pm$ 1 WR stars as well as a ratio of ${\rm N(WR)/N(O) = 0.012}$. Our estimations for both, the total number of WR stars and the WR stars over O-type stars ratio, are higher than the model predictions, mainly due to the large uncertainty in the estimation of WN stars.


\subsection{The gas accretion scenario}
\label{sec:gasacc}

From Figure \ref{fig:chemMaps} it is clear that the metallicity of the gas in CGCG 007-025 is not homogeneous across the galaxy. While the metallicity of the different star forming clumps is roughly the same (12 + log(O/H) $\sim$ 7.9), they stand out of the mean metallicity of the ISM ($\sim 8.3$). Remarkably, the SF clumps are at least 400 pc apart from the photometric centre of the galaxy. 

In closed-box chemical evolution models \citep{jr:BOOKtinsley1980} the available gas reservoir in the galaxy is contaminated by previous SF episodes, so that each new burst of star formation has higher metallicity than the previous one. Moreover, the gas will fall to the centre of the potential well, generating SF regions close to the centre of the galaxy. Accordingly, the metallicity will be higher where the stellar mass density is also higher, creating metallicity gradients that typically decrease inside out \citep[among others]{jr:vilchez1988,jr:vdkruit2011,jr:moran2012}. Thus, off-centred metallicity drops associated with star-forming regions cannot be explain within this scenario. 
Metallicity inhomogeneities can be generated via three different processes, that can operate simultaneosly: the mixing of metals within the ISM \citep{jr:werk2011}; strong outflows of metals \citep{jr:chisholm2015}; and inflows of metal-poor gas in tidally interacting systems \citep{jr:pearson2019} or via clumpy gas accretion from the cosmic web \citep[e.g., ][]{jr:sanchezalmeida2015, jr:ceverino2016}. Below we give a general view of the three scenarios, and expose how our findings favour a metal-poor gas accretion event.

\subsubsection*{Mixing of metals}

Metal production in galaxies traces star formation, and is highly concentrated toward the centres of galactic discs. In the absence of large-scale metal mixing, the density of metals should be directly proportional to the density of stars. However, several studies have found that metallicity gradients are flatter than density gradients \citep[e.g., ][]{jr:queyrel2012}. \citet{jr:petit2015} simulated an isolated disc galaxy and studied the effects of shear and turbulence in its metal distribution. They found the metal distribution is highly affected by both, with turbulence being the major agent in small scale structures and shear transferring momentum to turbulence from large scale to small scale dissipation.
In the absence of perturbations, such as bars or spiral arms, large scale inhomogeneities are dissipated slower than small scale ones mainly by the turbulence of the gas. \citet{jr:petit2015} also shown that, in a axisymmetric system, the mixing timescales are typically a fraction of the orbital period, i.e. a few hundred of Myr. 
Galactic fountains also redistribute the gas mass on large scale and so smear out the metallicity inhomogeneities \citep[e.g., ][]{jr:marasco2012, jr:gasaccretion}

Reconstruction of the star formation history for the underlying host galaxy (del Valle-Espinosa et al., in prep) indicates that star formation ceased over 1 Gyr ago. In the absence of any other mechanisms, metal mixing should produce a more homogeneous and flat oxygen abundance distribution than the one observed in the left panel of Figure \ref{fig:chemMaps}.

\begin{figure}    
    \begin{center}
        \includegraphics[trim=10 60 0 0, clip, width=0.48\textwidth]{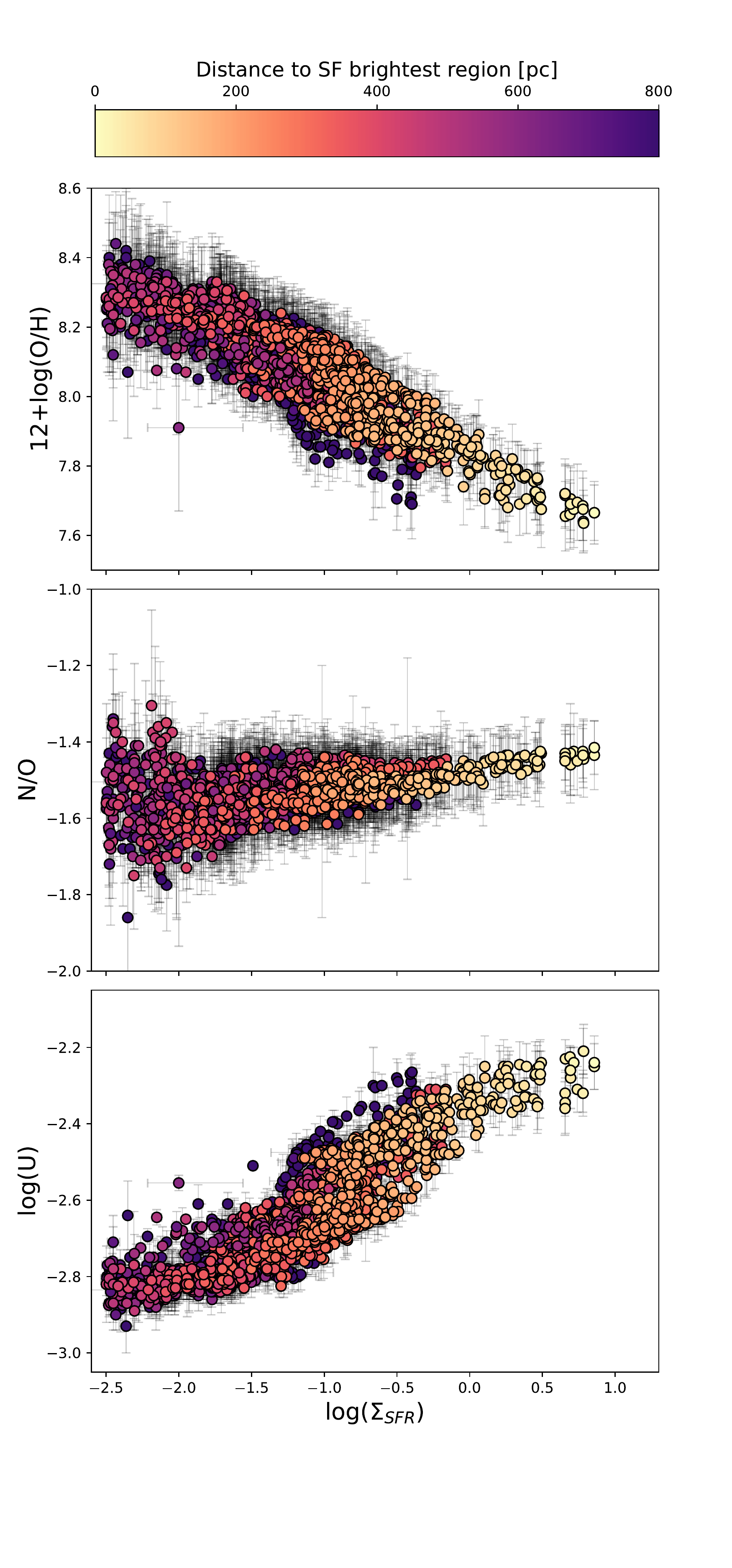}
    \end{center}
\caption{From top to bottom: oxygen abundance, nitrogen-to-oxygen and ionising parameter versus SFR surface density for every spectral bin. Markers are colour-coded by the distance to the centre of the brightest clump. While the oxygen abundance (ionising parameter) anti-correlates (correlates) with the SFR surface density, the nitrogen-to-oxygen shows no correlation with the $\Sigma_{\text{SFR}}$. }
\label{fig:SFROHNO}                 
\end{figure}

\subsubsection*{Outflow of metals}

The evolution of the most massive stars generates energetic winds and supernova explosions. These winds and blasts are the two main sources responsible of gas expulsion \citep{jr:negfeedbackI,jr:negfeedbackII}. In low-mass galaxies, the metal-enriched gas existing in the ISM can be easily swept out and returned to the IGM due to their shallow potential wells.

The optical tracers of the stellar winds and supernova explosions are broad emission lines. As mentioned in Section \ref{sec:methods}, and displayed in Figure \ref{fig:Halphamaps}, the H$\alpha$ emission in the brightest SF knot is described with a three-gaussian model. Our intermediate component around this region has an extension of 0.7$\times$0.4 kpc$^2$, with a median internal velocity dispersion of $\sim$200 km s$^{-1}$. 
\citet{jr:marasco_2022} studied in more detail the broad components of a sample of star forming dwarf galaxies which includes CGCG 007-025. Their analysis presents two different approaches for the treatment of the broad components.
Their Scenario 2 is more similar to our fitting procedure, and they found a wind component with an radial extension of 0.5 kpc and a velocity dispersion of 167 km s$^{-1}$. Their analysis, in alignment with other studies \citep{jr:lelli2014,jr:mcquinn2019} concludes that \textit{the baryonic feedback stimulates a gentle gas cycle rather than producing a large-scale blow out}. Ejection of metals from the centre of the galaxy will produce --anomalous-- enhanced O/H but also N/O ratios in the outer parts of the galaxy \citep{jr:belfiore2015,jr:luo_2021}. With the outflow characterisation by \citet{jr:marasco_2022}, and since the N/O distribution remains flat across the galaxy, we can rule out the outflow of metals as the main explanation for the metallicity inhomogeneities mapped. 

\subsubsection*{The accretion of metal-poor gas}

The simplest explanation for localised metallicity drops in a single system is the accretion of external metal-poor gas, induced by a recent merger, interaction or cosmological gas accretion. 
The same kind of metallicity deficit associated with bright star-forming regions has also been observed in several starbursting dwarf galaxies \citep{jr:tadpoles,jr:BCDinhomogeneities}. This behaviour is fairly common at high redshift \citep{jr:straughn2006,jr:windhorst2006,jr:elmegreen2007,jr:elmegreen2010}, and in the local universe is usually associated to low mass \citep[$M_{\star} < 10^{10.5} M_{\odot}$, e.g. ][]{jr:sanchezmenguiano2019} and metal-poor galaxies \citep{jr:papaderos2008,jr:ML2011,jr:filho2013}.

\citet{jr:localSFR-OH} studied the spaxel-to-spaxel anticorrelation between the gas-phase metallixity and the SFR in a sample of 14 local star-forming dwarf galaxies, expected if variable external metal-poor gas fuels the star-formation process. 
Similarly, in our Figure \ref{fig:SFROHNO} we show the oxygen abundance against the SFR surface density for all the analysed bins (upper panel). The quantities are clearly anti-correlated, but it is necessary to discard variations in the physical conditions of the gas to confirm this anticorrelation is triggered by metal-poor gas accretion. We also display the log(N/O) (log(U)) versus the SFR surface density in the middle (lower) panel of Figure \ref{fig:SFROHNO}. Changes in the nitrogen-to-oxygen ratio are linked with changes in the chemical composition of the gas, whereas the ionisation parameter changes with the age and mass of the stellar population. The N/O remains constant across the galaxy, while the log(U) correlates with the $\Sigma_{\text{SFR}}$. These three relations are naturally explained if a significant amount of low metallicity gas was accreted, reducing the gas-phase metallicity and enhancing star formation, but not altering the relative metal content of the ISM, including N/O.
Moreover, the underlying population of CGCG 007-025 is much older than the current SF burst (1 Gyr versus 5 Myr, respectively), leaning the scale to the gas accretion scenario rather than gas consumption as the main responsible of the recent SF episode.

\section{Conclusions}
\label{sec:conclusions}

Local starbursting galaxies present a unique environment to study star formation processes akin to those at higher redshift with much greater sensitivity and spatial resolution. In this paper we present a study of the chemodynamical properties of the ionised gas in CGCG 007-025. CGCG 007-025 is a dwarf spheroidal galaxy undergoing an off-centre starburst event--most likely as a result of an interaction with a nearby dwarf companion. Our main results, based on the analysis of the ionised gas present in the galaxy, are summarise below:

\begin{itemize}
    \item We derive the velocity maps for the different kinematic components of the H$\alpha$ line. The velocity field of the narrow component is highly perturbed, with no indication of rotation. The intermediate component presents a velocity profile compatible with a bipolar structure in expansion. The broad component, only detected in H$\alpha$ and in the brightest SF knot, displays a velocity distribution which resembles to a dusty expanding bubble, and could be originated in denser gas ($n_e \sim 10^6~{\rm cm^{-3}}$).
    \item Using \textsc{HII-CHI-mistry} \citep{pr:hiichim}, we obtain the chemical properties of the gas. We find a homogeneous distribution of N/O, whereas the distribution of oxygen abundance decreases at the position of the SF clumps. The ionisation parameter distribution presents the opposite behaviour.
    \item We compute line ratios between species with different ionisation potentials. The 2D map of [\ion{O}{iii}]/H$\beta$ ([\ion{S}{ii}]/H$\alpha$) peaks (decreases) at the position of the SF clumps. The [\ion{O}{iii}]/[\ion{S}{ii}] map reveals the extension of the highly ionised gas, expanding outside of the SF clumps and reaching the ISM. The position of the spaxels in the BPT diagrams reveal the ionisation is coming mainly from the newborn massive stars, with no hints of other mechanisms playing an important role in the galaxy.
    \item Thanks to the depth of the MUSE datacube, we are able to detect the WR red-bump in the brightest SF complex of the galaxy. The detection of the bump allows to make a better estimation of the age of this burst, 5.1 $\pm$ 1.2 Myr, as well as the number of WR stars present in it, 103 $\pm$ 26. 
    \item The overall relation between the distribution of metals and the $\Sigma_{\text{SFR}}$ suggest the accretion of metal poor gas to be responsible for the starbursts we are observing now in this galaxy.
\end{itemize}

This paper is devoted to the physical properties and origin of the ionised gas in CGCG 007-025. However, this galaxy presents an underlying old population as well as a NSC. The stellar properties of this galaxy will be discussed in a companion paper (del Valle-Espinosa et al., in prep.).

\section*{Acknowledgements}


We thank the anonymous referee for a constructive report that has improved the presentation of results. 
M.G.V.E acknowledges the support of the UK Science and Technology Facilities Council. 
V.F. acknowledges financial support provided by FONDECYT grant 3200473. 
R.A. acknowledges support from ANID Fondecyt Regular 1202007. 
BG-L acknowledges support from the Spanish Ministry of Science and Innovation through the Spanish State Research Agency (AEI-MCINN/10.13039/501100011033) through grants PID2019-107010GB-100 and the Severo Ochoa Program 2020-2023 (CEX2019-000920-S). 

\textit{Software}: this work made an extensive use of \textsc{Python}, and more specifically of \textsc{lmfit} \citep{pr:lmfit}, \textsc{numpy} \citep{pr:numpy}, \textsc{astropy} \citep{pr:astropy}, \textsc{scipy}, \textsc{matplotlib} \citep{pr:matplotlib} and \textsc{lineid\_plot} \citep{pr:lineid_plot}.

\section*{Data Availability}


Based on data obtained from the ESO Science Archive Facility with DOI \url{https://doi.org/10.18727/archive/41}. 



\bibliographystyle{mnras}
\bibliography{mnras} 




\appendix

\section{List of emission lines}

\begin{table}
    \centering
    \begin{tabular}{rcl} \hline
    Element & Wavelength & Components \\ 
    & \AA & \\ \hline
     [\ion{Ar}{iv}] & 4740.120 & N \\ 
     
     \ion{H}{ii} & 4861.333 & N+I \\ 
     
     \ion{He}{i} & 4921.931 & N \\
     
     [\ion{O}{iii}] & 4958.911 & N+I \\ 
     
     [\ion{Fe}{iii}] & 4987.250 & N \\ 
     
     [\ion{O}{iii}] & 5006.843 & N+I \\ 
     
     \ion{He}{i} & 5015.678 & N \\ 
     
     [\ion{O}{i}] & 6300.304 & N \\ 
     
     [\ion{S}{iii}] & 6312.060 & N \\ 
     
     [\ion{O}{i}] & 6363.776 & N \\ 
     
     [\ion{N}{ii}] & 6548.050 & N \\ 
     
     \ion{H}{ii} & 6562.819 & N+I+B \\ 
     
     [\ion{N}{ii}] & 6583.460 & N \\ 
     
     \ion{He}{i} & 6678.152 & N \\ 
     
     [\ion{S}{ii}] & 6716.440 & N+I \\ 
     
     [\ion{S}{ii}] & 6730.810 & N+I \\ 
     
    \end{tabular}
    \caption{List of emission lines modelled and their restframe wavelengths. \textit{N} stands for narrow profiles, \textit{I} for intermediate components and \textit{B} for broad profile.}
    \label{tab:emlist}
\end{table}


\bsp	
\label{lastpage}
\end{document}